\documentclass[journal]{IEEEtran}

\usepackage{CJK}
\usepackage{indentfirst}
\usepackage{graphicx}
\usepackage{float}
\usepackage{dirtree}
\usepackage{amsmath,amssymb}
\usepackage{setspace}
\usepackage{mathtools}
\usepackage{float}
\usepackage[normalem]{ulem}
\usepackage{multirow}
\usepackage{array}
\usepackage[ruled]{algorithm2e}
\usepackage{amsmath}
\usepackage{multicol}
\usepackage{multirow}
\usepackage{cite}
\newcommand{\li}{\uline{\hspace{0.5em}}}

\begin{document}
\let\sss= \scriptscriptstyle
\let\s= \scriptstyle
\let\ttt= \textstyle

\title{Stylize Aesthetic QR Code}

\author{ Mingliang Xu,~Hao Su,~Yafei Li,~Xi Li,~Jing Liao,~Jianwei Niu,~Pei Lv and Bing Zhou

\thanks{The corresponding author is Dr. Yafei Li. (e-mail: ieyfli@zzu.edu.cn.)}
\thanks{Mingliang Xu, Hao Su, Yafei Li, Pei Lv, Bing Zhou are with Zhengzhou University. Li Xi is with Zhejiang University, Jing Liao is with Microsoft Research Lab-Asia, Jianwei Niu is with BeiHang University.}
}

\markboth{IEEE TRANSACTIONS ON MULTIMEDIA,~Vol.~14, No.~8, August~2017}%
{Shell \MakeLowercase{\textit{et al.}}: Bare Demo of IEEEtran.cls for IEEE Journals}

\maketitle

\begin{abstract}
With the continued proliferation of smart mobile devices, \emph{Quick Response (QR)} code has become one of the most-used types of two-dimensional code in the world. Aiming at beautifying the visual-unpleasant appearance of QR codes, existing works have developed a series of techniques. However, these works still leave much to be desired, such as personalization, artistry, and robustness. To address these issues, in this paper, we propose a novel type of aesthetic QR codes, \emph{SEE (Stylize aEsthEtic) QR code}, and a three-stage approach to automatically produce such robust style-oriented codes. Specifically, in the first stage, we propose a method to generate an optimized baseline aesthetic QR code, which reduces the visual contrast between the noise-like black/white modules and the blended image. In the second stage, to obtain art style QR code, we tailor an appropriate neural style transformation network to endow the baseline aesthetic QR code with artistic elements. In the third stage, we design an error-correction mechanism by balancing two competing terms, visual quality and readability, to ensure the performance robust. Extensive experiments demonstrate that \emph{SEE QR code} has high quality in terms of both visual appearance and robustness, and also offers a greater variety of personalized choices to users.

\end{abstract}

\begin{IEEEkeywords}
QR code, style-oriented, visual aesthetics, robust.
\end{IEEEkeywords}

\IEEEpeerreviewmaketitle

\setlength{\parindent}{2em}

\section{Introduction}
\IEEEPARstart{W}{ith} the continued proliferation of the internet and smart mobile devices, \emph{Quick Response (QR)} code has become one of the most widely used information carriers in the world. However, the ordinary QR codes have visual-unpleasant appearances and consist of  monotonic black/white square \emph{modules} which are meaningless to human vision. Hence, the visual optimization of QR code has attracted extensive attentions from academia and industry.

 As shown in Fig. 1, existing works can be categorized into four types: i) \emph{embedded-type}~\cite{R1,MeasureQR,Lin2013Artistic,Garateguy2014QR,40Exemple} which embeds small icons utilizing the correction capability of QR codes; ii) \emph{deformation-type}~\cite{40Exemple} that changes the shape and color of the modules in QR codes, e.g. turning the square modules into round, triangle, star; iii) \emph{manual type}~\cite{40Exemple}~which is produced by manual design and rendering; iv) \emph{blended-type}~\cite{VS,HF,EF,Masic,TS,LQF,Kuribayashi2017Aesthetic,Fang2014An}~which blends a large image into QR code. Among these, the \emph{blended-type} is considered as the most promising technique for generating QR codes with the highest visual quality.
\begin{figure}[tb]
\centering
\includegraphics[width=2.8in]{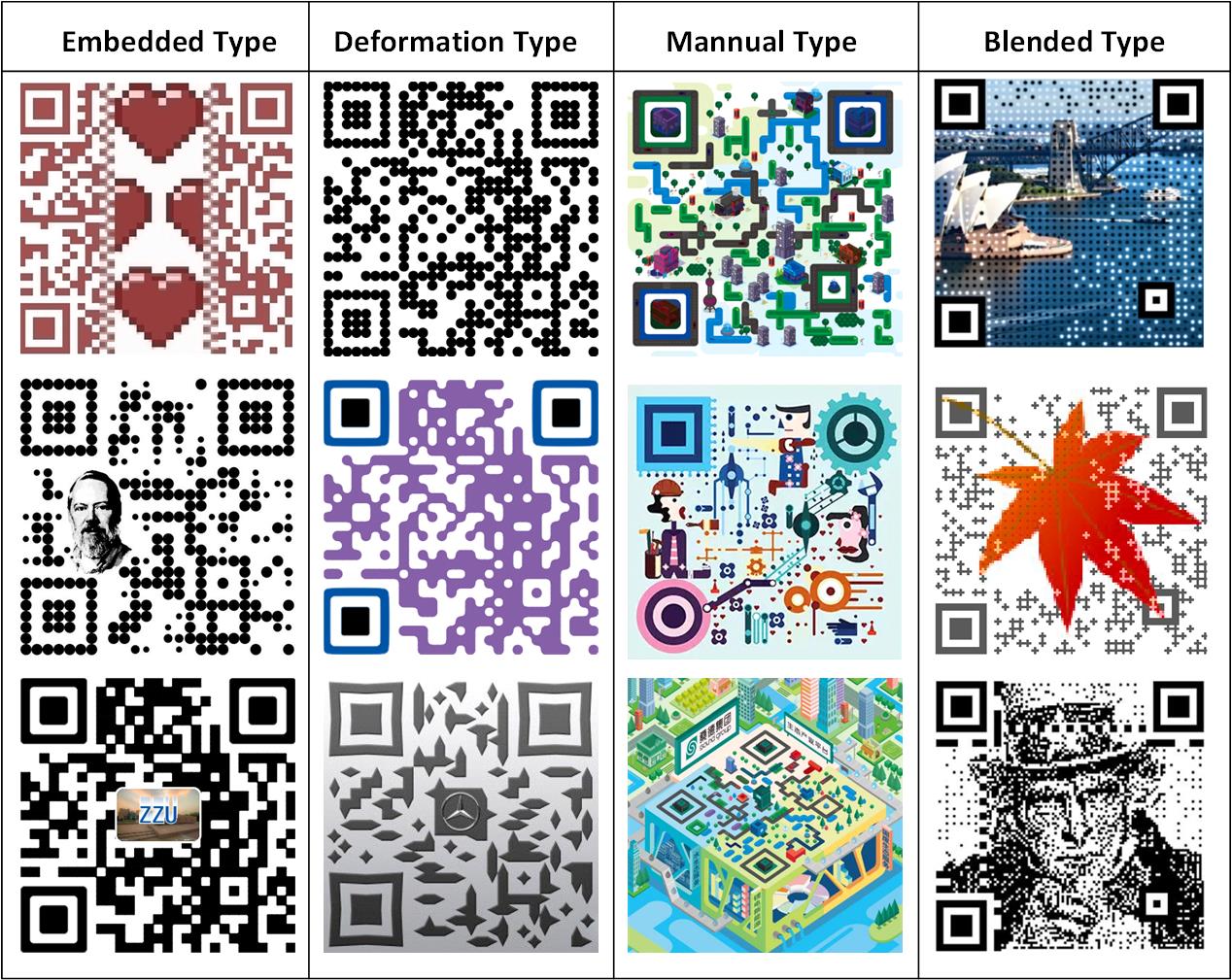}
\caption{Four types of existing approaches to beautify the QR codes.}
\end{figure}
\begin{figure}[tb]
\centering
\includegraphics[width=2.8 in]{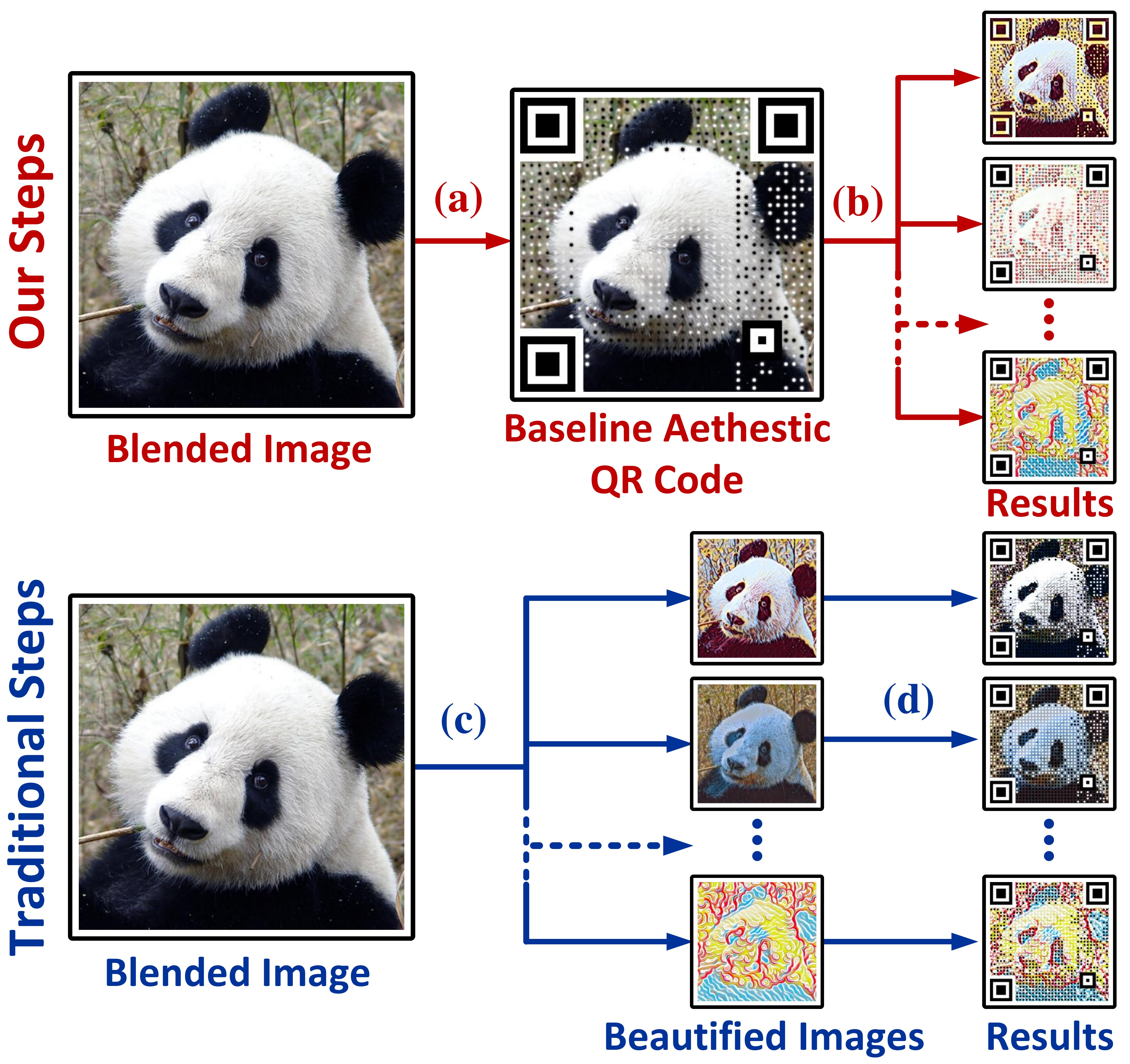}
\caption{(a)(b) Our approach directly beautifies the baseline aesthetic QR code to generate art style results. (c)(d) If traditional blended-type methods desire to produce art style results, they must combine the beautified images and the invariable black/white encoding modules. In fact, their results are still baseline aesthetic QR codes which only blend with artistic images  (cf. Fig. \ref{fig:1Introduction-moudle}(a)-(d)). Compared with steps (c)(d), our approach has two merits: i) We simplify the steps to generate. ii) We further transfer the blended image and modules into a unified style while enhancing their visual appeal (cf. Fig.\ref{fig:1Introduction-moudle} (e)-(l)). }
\label{fig:1Introduction-XiongMao}
\end{figure}
\begin{figure*}[t]
\centering
\includegraphics[width=6.4in]{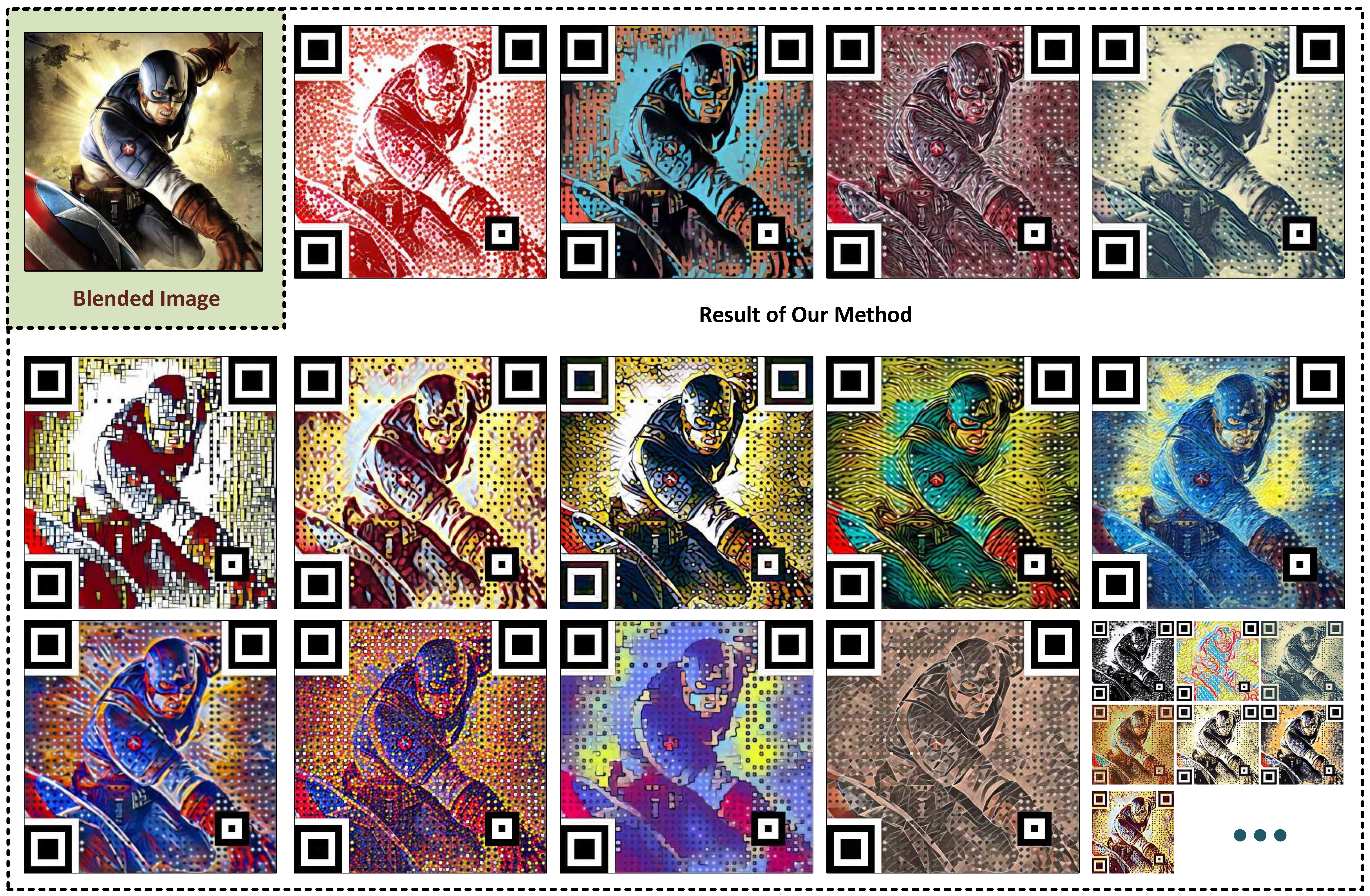}
\caption{Some examples of our SEE QR code, which are artwork-like. Moreover, users can produce them in various visual-pleasant artistic styles via a single blended image.}
\label{fig:1Introduction-Meiguoduizhang}
\end{figure*}
\begin{figure}[t]
\centering
\includegraphics[width=3.3 in]{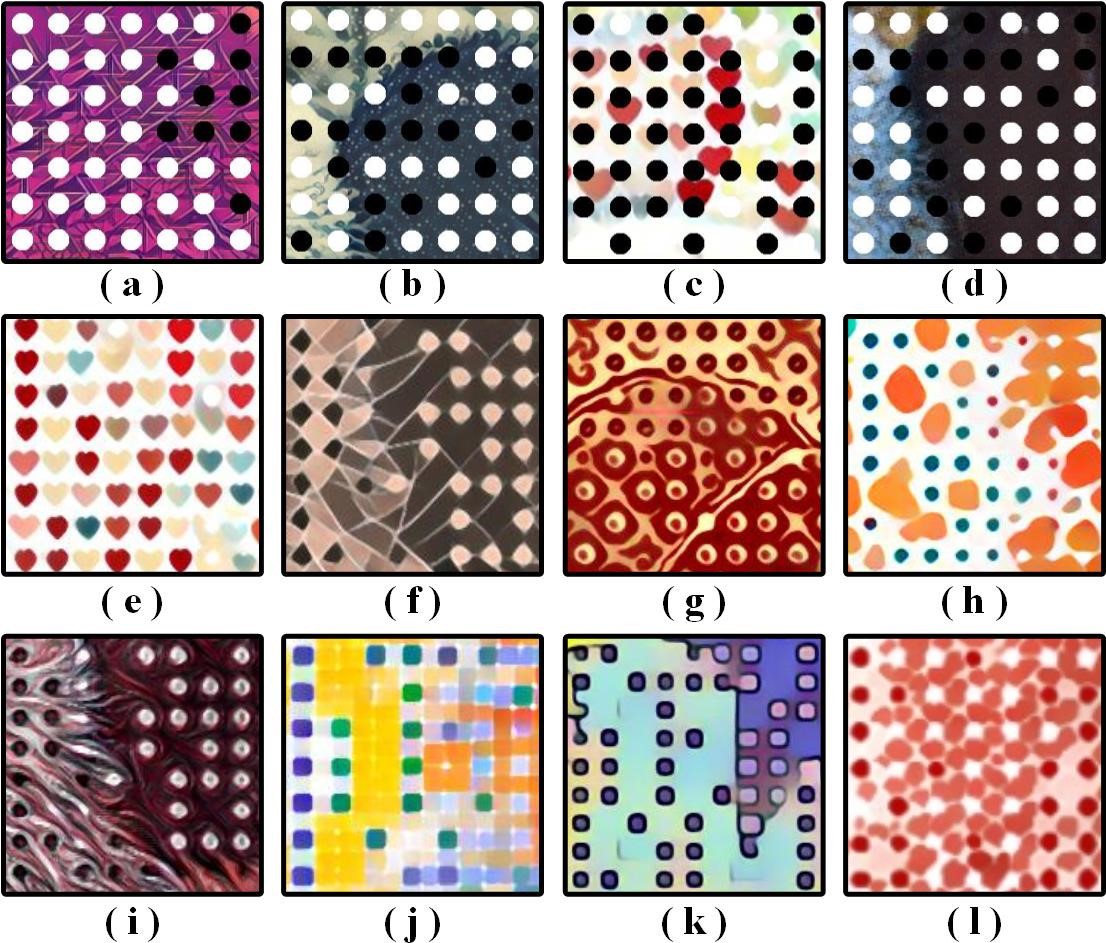}
\caption{(a)-(d) Encoding modules in the results of traditional steps (cf. Fig. 2), which have noise-like invariable appearances. (e)-(l) Encoding modules in our SEE QR code (cf. Fig. 2), where the blended images and modules are simultaneously endowed with attractive artistic elements in a unified style. }
\label{fig:1Introduction-moudle}
\end{figure}

Although existing works have improved the visual quality of QR code to some extent, they still leave much to be desired in terms of the following three aspects:
i)~\emph{Personalization}, mainstream works generate aesthetic QR codes with different appearances via changing the blended images. In fact, users always expect to produce personalized aesthetic QR codes in different styles by a unique blended image (e.g. logo, personal photo or trademark).
ii)~\emph{Artistry}, most existing works define ``beauty" is ``more similar to the blended image", they produce QR codes by combining an image with black/white modules directly, which lacks additional aesthetic refinement. Moreover, their resultant modules are always invariable and mechanical even blended with beautified images (cf. Fig. \ref{fig:1Introduction-moudle}(a)-(d)).
iii)~\emph{Robustness}, current works always utilize parameters to control the robustness of modules in a certain range, yet, without evaluating whether there exist errors in outputs. Therefore, partial error modules may decline the correction capability of encoding. Especially, in complex real scenarios, few modules smeared may result in unreadable due to the fault-tolerance limitation of QR codes.

Solving any of these issues without compromising other properties is a big challenge. In this paper, we propose an effective approach to  automatically produce robust art style QR code, called \emph{SEE (Stylize aEsthEtic) QR code}, by leveraging the CNN-based style transformation network. As shown in Fig. 3, SEE QR codes are style-oriented aesthetic codes that can be produced with various personalization styles by blending a single image. Moreover, our approach directly stylizes the baseline aesthetic QR code (cf. Fig. 2(a)(b)), which endows the encoding modules and blended image with unified artistic elements and enhances their visual appeal (cf. Fig. 4(e)-(l)). Finally, we design an error-correction mechanism by balancing two competing items, visual quality and readability, to ensure the robustness of SEE QR codes.

To summarize, our main contribution in this paper is fourfold:
\begin{itemize}
	\item We propose a new type of aesthetic QR codes, SEE QR code, which is personalized, artistical, and robust.
	\item We design an efficient algorithm for scheduling changeable modules in baseline aesthetic QR codes, which minimizes the visual contrast between black/white encoding modules and the blended image.
	\item We adapt a style transfer network for stylizing the baseline aesthetic QR codes, which effectively avoids the visual affecting of noise-like modules while reduces the encoding message loss during the transformation.
	\item We present an error-correction mechanism to ensure the robustness of each module in resultant QR codes by balancing two competing items, visual quality and readability.
\end{itemize}

\section{Related Work}
In this section, we review techniques related to our work which mainly refer to two topics, aesthetic QR code and style transfer.
\subsection{Aesthetic QR Code}
As elaborated in Section I, up to now, the {manual-type} techniques are high-cost and non-automatic, the {embedded-type} techniques and {deformation-type} techniques have unideal visual effect. In contrast, the {blended-type} techniques with good visual quality are the most promising approach that deserves further study. The details about representative existing blended-type works are as follows.

\begin{figure}
\centering
\includegraphics[width=3.5 in]{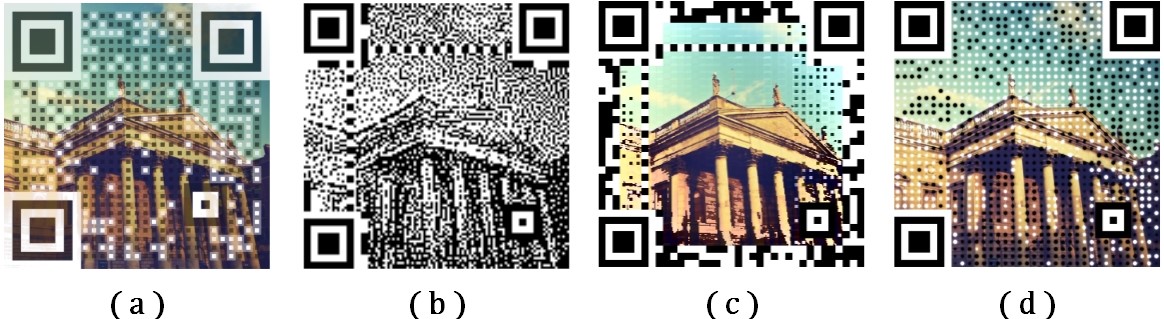}
\caption{Representative {blended-type} existing works on aesthetic QR codes: (a) Visualead QR code \cite{VS}, (b) Halftone QR code \cite{HF}, (c) Efficient QR Code \cite{EF}, (d) Two-Stage based QR code \cite{TS}.}
\label{fig:1Introduction-existing-method}
\end{figure}

Peled \emph{et al.} \cite{VS} developed a visual QR code generator called \emph{Visualead} (cf. Fig. \ref{fig:1Introduction-existing-method} (a)) which retains the original contrast between the encoding modules and the blended image to synthesize the aesthetic QR codes. However, the QR codes generated by the \emph{Visualead} have serious artifacts that notably reduce the visual content of the blended image.

Inspired by the technique of halftone, Chu \emph{et al.} \cite{HF} presented a novel style aesthetic QR code called halftone QR codes (cf. Fig. \ref{fig:1Introduction-existing-method}(b)). The idea of generating halftone QR codes is that they subdivide each module of the standard QR code into $3\times3$ submodules and bind the module's color to the center submodules while the remaining $8$ submodules are modified to balance the reliability and regularization. However, halftone QR codes are restricted by the substitution principle and only composed of black/white colors, which is still improvable in visual quality.

Aiming at blending image to a full-size area of the QR code, Lin \emph{et al.} \cite{EF} synthesize aesthetic QR code based on the Gauss-Jordan elimination used in the QArt method \cite{Cox} and improve the visual quality by a rendering mechanism, which is combined by the techniques of embedded-type and blended-type. However, such QR code is only suitable for the blending image of which saliency content is in the center of the image and not near the edges (cf. Fig. \ref{fig:1Introduction-existing-method}(c)).

Leveraging QArt method \cite{Cox}, Zhang \emph{et al.} \cite{TS} relocated the modules of QR code that depend on the visual saliency and edge features extracted from blended image (cf. Fig.\ref{fig:1Introduction-existing-method} (d)). This approach tends to distribute the black/white modules in the visual focus area, and output QR codes with visual-pleasant. However, this approach lacks an error correction mechanism to ensure the readability, and it is still improvable in the perspective of visual quality by adopting the image's global feature.

\vspace* {-10pt}
\subsection{Style Transfer}
\begin{figure}[b]
	\centering
	\includegraphics[width=3.5in]{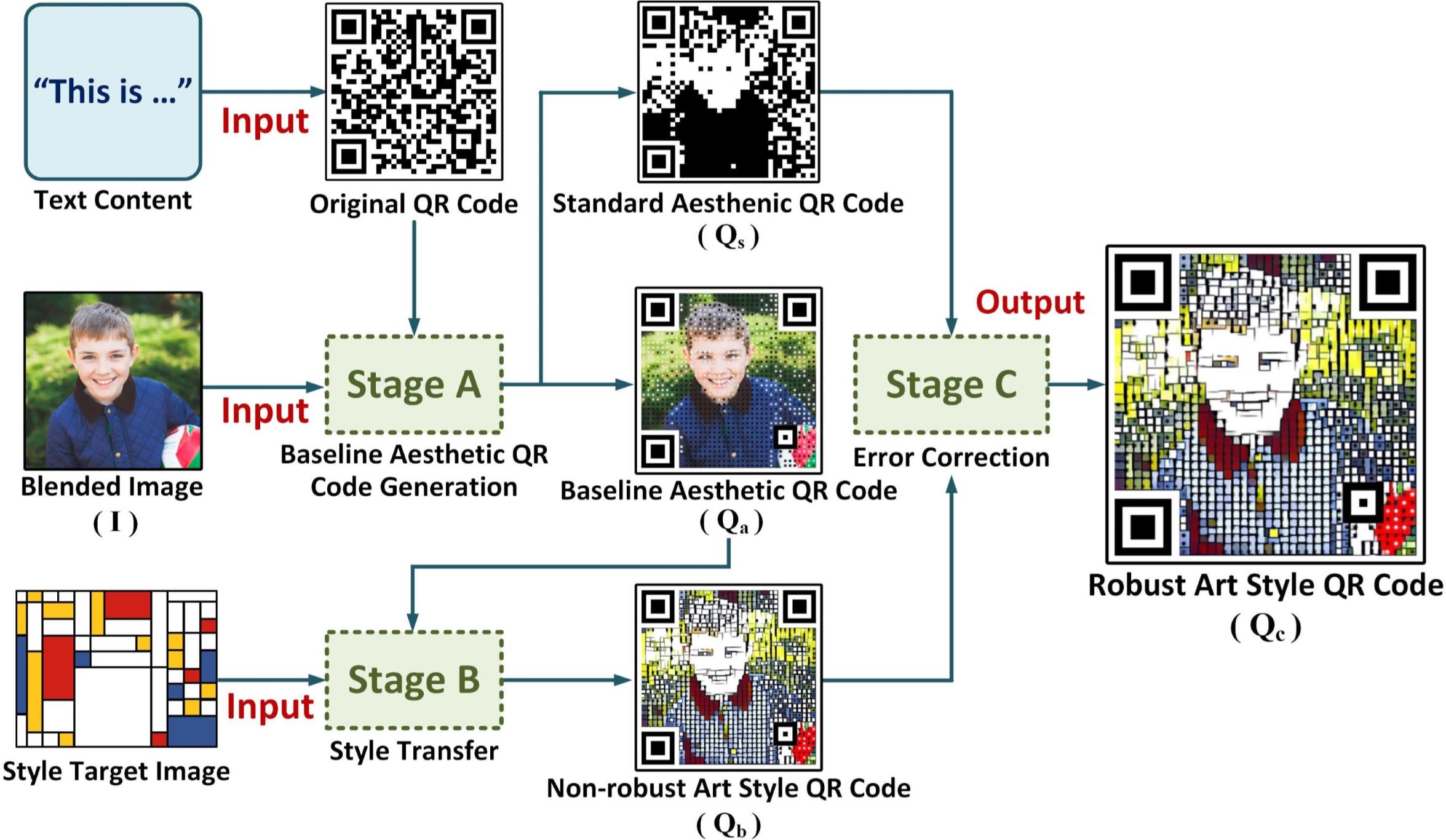}
	\caption{Overview of our approach, which consists of three stages, Stage A: baseline aesthetic QR code generation, Stage B: style transfer, and Stage C: error correction.}
	\label{fig:2Mehod}
\end{figure}
\begin{figure*}[t]
\centering
\includegraphics[width=6.5in]{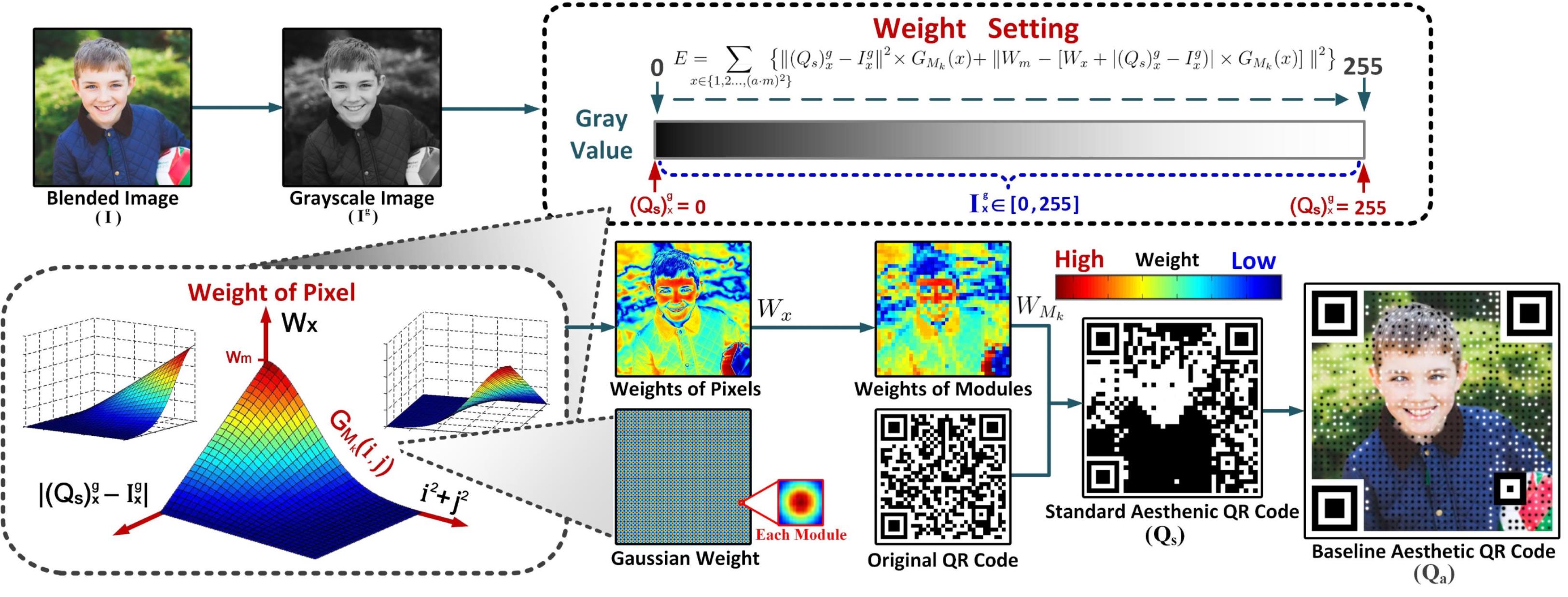}
\caption{Flowchart of Stage A. We calculate the priority weights of scheduling changeable modules depend on the gray value distribution of the blended image. This method can minimize the visual contrast between the blended image and noise-like black/white modules, which end up with a baseline aesthetic QR code $Q_{a}$.}
\label{fig:3PA}
\end{figure*}
Recently, style transfer has become a hot research topic in AI field, which is very related to texture synthesis. It can be interpreted as migrate artistic style from a style target image and blended with semantic information of the content target image.

Early studies on style transfer can be divided into two primary types: one type is based on optimization~\cite{Gatys,Li}, it produces impressive stylized image but is too time consuming for iterative optimizing; the other type is based on feed-forward network~\cite{Lff,Texture}, feed-forward generator network is trained for each specific style target image, the original time-consuming iterative optimization is replaced by a forward pass mechanism. Moreover, the learned style transfer feed-forward networks can output the results nearly real-time.

After that, Chen \emph{et al}. \cite{StyleBank} proposed an efficient method named StyleBank that allows a single network to simultaneously learn numerous styles. StyleBank is composed of multiple convolution filter banks, each filter bank explicitly represents one style for neural image style transfer. Based on this mechanism, style transfer can be realized through using StyleBank and auto-encoder.

Liao \emph{et al.} \cite{dia} proposed a novel technique called deep image analogy for visual attribute transfer. They combine the techniques of image correspondence and neural style transfer, and achieves prominent visual effect through establishing a pixel-level correspondence between two images which have similar semantic structure and different appearances.

In this paper, we attempt to solve the weaknesses stated above by combining the techniques of aesthetic QR code and style transfer to synthesize a novel type of robust art style QR codes with attractive appearances.

\section{Overview of our Approach}
As shown in Fig. \ref{fig:2Mehod}, our approach is consist of three stages denoted as Stage A, Stage B, and Stage C. Initially, depending on a novel strategy of scheduling changeable modules, we produce an optimized baseline aesthetic QR code $Q_a$, in Stage A. Then in Stage B, to endow $Q_a$ with artistic elements while considering the particularity of QR codes, we adapt a neural style transfer network for styling $Q_a$ and obtain an art style QR code $Q_b$. Finally, in Stage C, aiming at eradicating error modules in $Q_b$ to ensure the readability, an iterative-update based error correction mechanism is presented for outputting a robust art style result $Q_c$.

The details of Stage A, Stage B, and Stage C will be introduced in the following three Sections respectively. Table I summarizes the notations used throughout this paper.
\renewcommand\arraystretch{1.3}
\begin{table}[t] \small
	\caption{Summary of Notations}
	\centering
	\begin{tabular}{p{0.7cm}|p{7cm}}
		\hline
		\textbf{Name} & \textbf{Description}   \\
		\hline
		$ \ I$   &  The input image used for blending. \\
		$Q_{s}$  & The standard QR code produced in Stage A.\\
		$Q_{a}$  & The baseline aesthetic QR code produced in Stage A. \\
		$Q_{b}$  & The non-robust art style QR code produced in Stage B. \\
		$Q_{c}$  & The robust art style QR code produced in Stage C, namely SEE QR code.  \\
		$Q^{g}$  & The grayscale image of QR code $Q$. \\
		$Q^{b}$& The binary image of QR code $Q$. \\
		$Q^{c}$  & The color image of QR code $Q$. \\
		$Q^{t}$& The threshold of QR code $Q$. \\
        $Q_{x}$& The $x$-th pixel of QR code $Q$. \\
		$M_k$& The $k$-th encoding module of QR code. \\
		$S_{(k,r)}$& The $k$-th circular encoding spot of radius $r$ and concentric with $M_k$.   \\
		\hline
	\end{tabular}
	\label{table:meaning}
\end{table}
\renewcommand\arraystretch{1}
\begin{figure*}[t]
	\centering
	\includegraphics[width=6in]{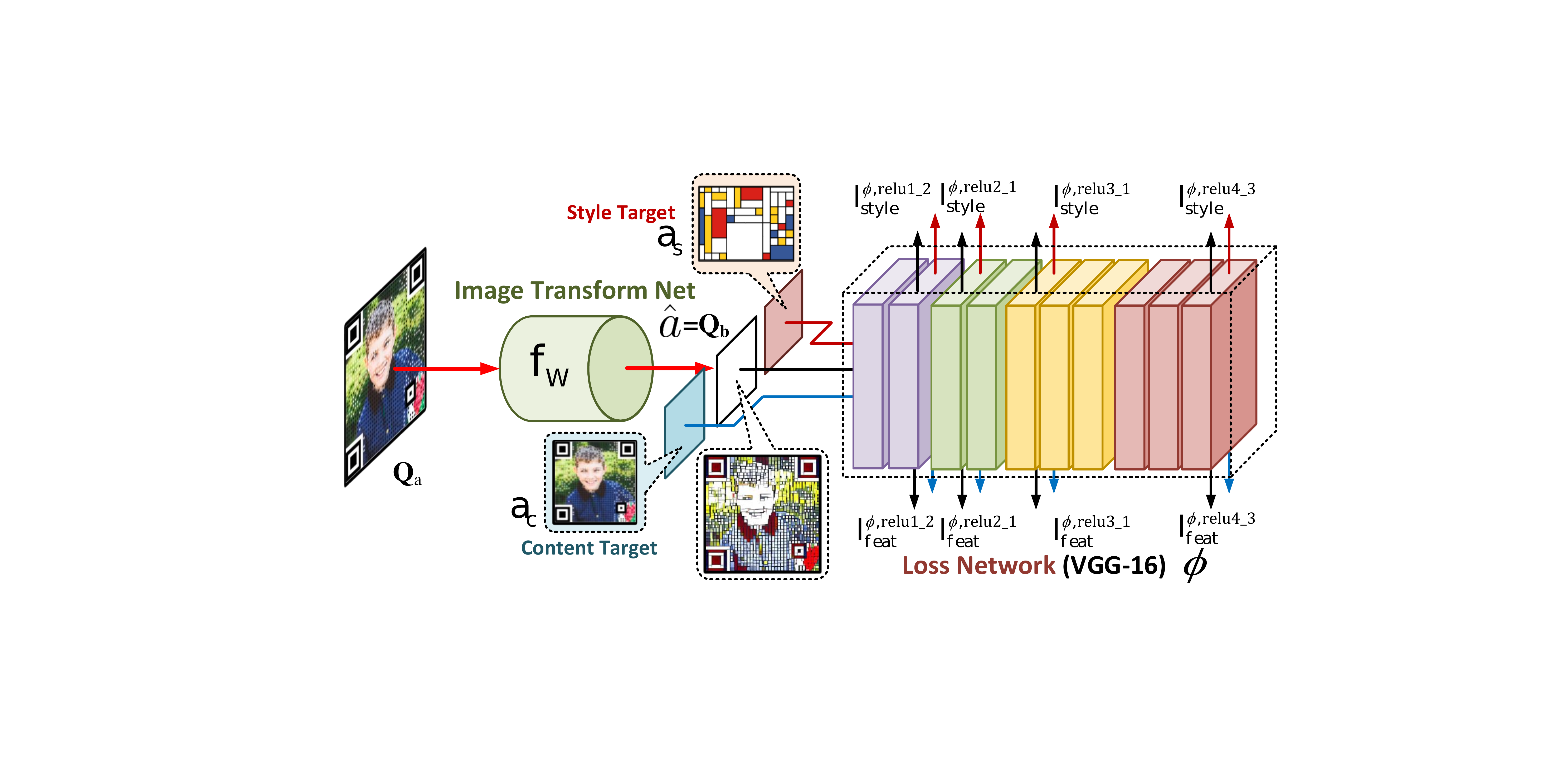}
	\caption{Flowchart of Stage B. We roughly follow the style transfer system proposed by \cite{Lff}. Aiming at enhancing the adaptability for stylizing baseline aesthetic QR codes which have dense black/white modules, we further adapt the layers of reconstructing style/content features loss in the loss network $\boldsymbol{\phi}$, and finally produce non-robust art style QR code $Q_b$.  } 
	\label{fig:3PB}
\end{figure*}
\section{Stage A:~Aesthetic QR Code Generation}
\subsection{Basic of Stage A}
QR code is based on the coding rules of Reed-Solomon (RS) code and expressed as square encoding modules. Cox \cite{Cox} proved that \emph{Gauss-Jordan elimination procedure} can be employed to schedule changeable modules in a limited range without compromising the machine readability.

Mainstream works (e.g. \cite{HF,EF,Masic,TS,LQF}) always manipulate the QArt method \cite{Cox} to schedule the changeable modules by considering local visual features of the blended image, such as saliency map, edge map, or ROI (region of interesting). Unlike them, in Stage A, we propose an effective strategy that depends on the global gray values of blended image $I$ and produces a baseline aesthetic QR code $Q_a$, which minimizes the visual contrast between $I$ and the noise-like black/white modules.

In the grayscale blended image $I^g$, the gray value of each pixel is in $[0, 255]$ while that of the black and white modules are constant $0$ and $255$ respectively. Accordingly, the visual contrast is minimized when the gray value of module is most approximate to that of the corresponding pixels in $I^g$. In other words, when the pixels in $I$ with the darkest/lightest color are preferentially scheduled with black/white module, the visual performance of $Q_a$ will be improved a lot.
\subsection{Generating process}
In this paper, we utilize the QR code of version 5 and error correction level $L$ as the default setting. We first generate a grayscale copy $I^g$ of $I$, and divide $I^g$ into $m\times m$ modules of size $a\times a$ pixels, which adheres to the ISO standard \cite{ISO}. $W_{M_k}$ denotes the normalized priority weight of scheduling module $M_k$, which is defined as
\vspace{0.3cm}
\begin{equation}
W_{M_k}= \frac{1}{W_m}  \sum_{x\in M_k} W_{x} \ ,
\label{equation:2PA-W_Mk}
\vspace{0.3cm}
\end{equation}
where $x$ is a pixel of module $M_k$, $W_x$ is the weight of $x$ with a maximum value of $W_m$, for the convenience of calculation, here $W_m=\;$255. We assign all pixels with different weights $W_x$ and minimize the following energy function to automatically calculate the best $W_x$ for each pixel
\begin{equation}
\begin{aligned}
E= &  \!\!\!\!\!\!\!\! \sum_{ {x\in \{1,2...,(a\cdot m)^2  \}}^{\vphantom{'}} } \!\!\!\!\! \!\!\! \left\{  \left \| (Q_s)_x^g-I_x^g \right \|^2  \! \times G_{M_k}(x)  +   \right. \left \| W_m   \right. \\
   & \left.  \left.  \  - \; \; [ W_x+| (Q_s)_x^g-I_x^g| \times G_{M_k}(x) ]  \;  \right \|^2   \right\}  \ ,
\end{aligned}
\label{equation:Set Weight}
\end{equation}
where $(Q_s)_x^g$ is 0 or 255. The first term in the summation ensures that the gray value of each pixel in the resultant standard QR code $Q_s$ should be similar to $I^g$. The second term in the summation ensures that pixels with smaller gray value differences are assigned higher weights. $G_{M_k}(x)$ denotes a \emph{Gaussian weight function} we defined, due to the rule that the pixels closer to the center with higher probability to be sampled during scanning, $\sum_{x \in M_k}G_{M_k}(x)=1$. Here,
\begin{equation}
\label{equation:PA-Gauss}
G_{M_k}(x)=G_{M_k}(i,j)=\frac{1}{2\pi \sigma^2}e^{-\frac{i^2 + j^2}{2\sigma^2}},
\end{equation}
where $i$, $j$ respectively denote the horizontal-ordinate and vertical-ordinate of pixel $x$ when setting up a coordinate system with an origin at the center of $M_k$, $\sigma=\frac{a-1}{6}$, $a$ is the side length of $M_k$.

According to Eq.(\ref{equation:2PA-W_Mk}) and Eq.(\ref{equation:Set Weight}), the weight of each module in $I^g$ is calculated to obtain an $m\times m$ weight matrix $W$. For the $k$-th module with a bigger weight, we give higher priority to it for being scheduled a module which has the same binary result as the $k$-th module of $I^g$, that is
\begin{equation}
\begin{aligned}
(Q_s)^b_{M_k}=Round \left\{ \left[ \sum\nolimits_{ x\in M_k} I_x^g \cdot G_{M_k}(x) \right] \times \frac{1}{255}  \right\},
\end{aligned}
\label{equation:Ig_and_Qs}
\end{equation}
where $(Q_s)^b_{M_k}$ is 0 or 1. Afterward, according to $W$, we manipulate \emph{Gauss-Jordan elimination procedure} mentioned in \cite{Cox} to schedule changeable modules in original QR code and produce the module-based standard QR code $Q_s$.

Finally, we replace the black/white square modules $M_k$ of $Q_s$ with circular spots $S_{(k,r)}$ of radius $\frac{1}{4}a$ (the setting of $S_{(k,r)}$ will be detailed in subsection B, Section VI) and fill the rest area with the corresponding pixels of $I$, thus the baseline aesthetic QR code $Q_{a}$ is produced.
\begin{figure}[t]
\centering
\includegraphics[width=3in]{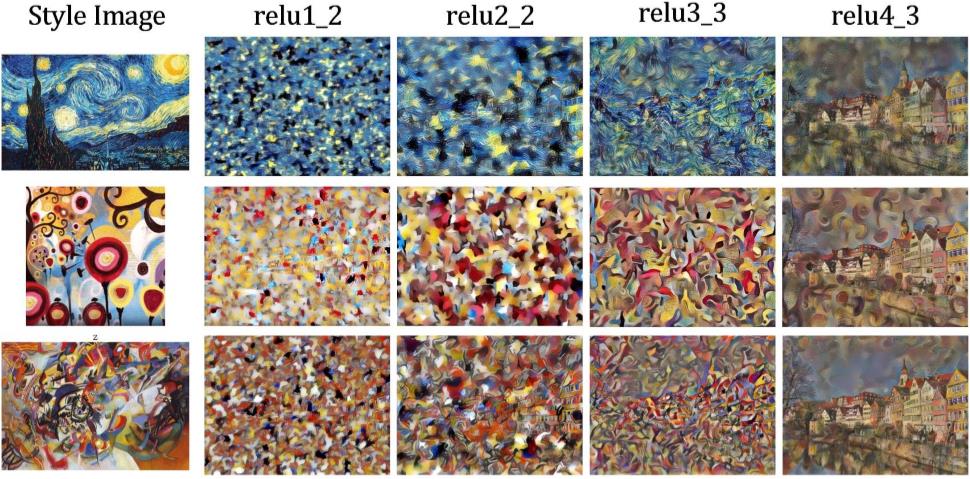}
\caption{We evaluate the style features reconstructed from the pretrained VGG-16 loss network in layers relu1$\li$2, relu2$\li$2, relu3$\li$3 and relu4$\li$3. Meanwhile, we find that the features reconstructed from low-level layers extremely similar to the dense encoding modules of aesthetic QR codes, which inspired us.}
\label{fig:3PB-Gfig:3PB-Gecengduibiecengduibi}
\end{figure}
\section{Stage B:~Style Transfer}
\setcounter{subsubsection}{0}
\subsection{Framework of style transfer system}
Fig. 8 shows our system framework of the style transfer that roughly follows the architecture proposed in [16]. It is composed of two primary parts: a deep residual convolutional neural network $f_W$ and a pretrained loss network $\phi$.

In this system, $f_W$ is used for transforming image $Q_a$ into image $Q_b$ via the mapping $\hat{a}=f_W(x)$, which is trained by stochastic gradient descent to minimize a weighted combination of loss functions
\begin{equation}
W^{*}= \mathop{\arg\min}_{ \quad  W  } E_{x,\{a_i\}} \left[ \ \sum_{i=1}\lambda_i l_i (f_W (x),y_i ) \ \right].
\end{equation}

${\phi}$ is the 16-layer VGG network \cite{VGG16} pretrained on ImageNet \cite{ImageNet}, which used for image classification to measure perceptual differences in the perspectives of content and style between the output images $\widehat{a}$ and a target image $a_i$. The loss function contains a feature reconstruction loss $l_{feat}^{\boldsymbol{\phi}}$ and a style reconstruction lose $l_{style}^{\boldsymbol{\phi}}$.

\begin{equation}
\begin{aligned}
  l_{feat}^{\phi,j}(\widehat{a},a)&=\frac{1}{C_jH_jW_j}\| \phi_j(\widehat{a})-\phi_j(a)\|_2^2
\\l_{style}^{\phi,j}(\widehat{a},a)&=\| G_j(\widehat{a})-G_j(a)   \|_F^2  \qquad \quad
\end{aligned},
\end{equation}
Once the training process of $f_W$ is finished, we can use $f_W$ to transform the input images $Q_a$ into stylized results $Q_b$ in real-time.

\subsection{Adaptability adjustment of network}
In Stage B, when the baseline aesthetic QR code $Q_a$ with dense black/white encoding modules are treated as the content target of the style transfer, two key issues for consideration: i) For the robustness, the loss of encoding messages incurred by style transformation should be minimized; ii) For the visual quality, avoiding the visual damage caused by the noise-like modules is necessary. Aiming at these goals, we carry out the following tasks.

Initially, as illustrated in Fig.~\ref{fig:3PB-Gfig:3PB-Gecengduibiecengduibi}, we evaluate the style features reconstructed from layers relu1$\li$2, relu2$\li$1, relu2$\li$2, relu3$\li$1, relu3$\li$2, relu3$\li$3, relu4$\li$1, relu4$\li$2, relu4$\li$3 of the pre-trained VGG-16 loss network $\boldsymbol{\phi}$. Afterward, we find that in lower layers (e.g., relu1$\li$2 and relu2$\li$2), the reconstructed style features put more emphasis on localization that presents as discrete textures and dense fragments, which is extremely similar to the encoding modules of $Q_a$. Correspondingly, with the increasing of the layers in VGG-16, the reconstructed features are gradually changed from localization to globalization. These phenomena inspired us.
\renewcommand\arraystretch{1.5}
 \begin{table}[t]\footnotesize
\caption{The refinment for features reconstruction layers}
\centering
\begin{tabular}{p{1.3cm}|p{6cm}}
\hline
\hline
\multicolumn{2}{c}{ \ \ \textbf{Layers of Style Feature Reconstruction Loss}}   \\
\hline
 \ [16] & \ \  \qquad  relu1$\li$2, relu2$\li$2, relu3$\li$3, relu4$\li$3    \\
\hline
\ Ours  & \ \  \qquad relu1$\li$2, relu2$\li$1, relu3$\li$1, relu4$\li$3 \\
\hline
\multicolumn{2}{c}{ \ \ \  \ \; \textbf{Layers of Content Feature Reconstruction Loss}}    \\
\hline
\ [16]& \ \  \qquad relu3$\li$3 \\
\hline
\
Ours  & \ \  \qquad relu1$\li$2, relu2$\li$1, relu3$\li$1, relu4$\li$3 \\
\hline
\hline
\end{tabular}
\renewcommand\arraystretch{1}
\label{table:layer-modify}
\end{table}

\renewcommand{\thefootnote}{\fnsymbol{footnote}}

 Consider the special characteristics of $Q_a$ (i.e., $Q_a$ has discrete, fragmented, and dense encoding modules), as shown in Table \ref{table:layer-modify}, we pertinently adapt the loss reconstruction layers of the style/content features. As expected, after the modification, we obtain desired results\footnote{The experimental results are detailed in subsection B, Section VII.} on both visualization and robustness.

\section{Stage C:~Error Correction}
 In Stage B, we have significantly reduced the loss of encoding messages. However, there may still exist few error modules in $Q_b$. Thus, we present an error-correction mechanism in this stage to detect and correct these errors by balancing the robustness and visual quality, which leads to a robust art style result $Q_{c}$.
\subsection{Basic of QR Code decoding}
QR code is based on the rules of RS code, which cannot be decoded once sampled errors data exceed the correction capability. To address this issue, we divide the translating process of QR code messages into two steps: \emph{sampling} and \emph{thresholding}. Note that a valid decoding of QR codes requires these two steps both correct.

For \emph{sampling}, we follow the rules in ZXing~\cite{ZXing} which is the most widely used library for QR code codec. ZXing rules that the sampled encoding message is only related to each module's center pixel. Following this rule, the sampled center pixels of all modules are further grayed and thresholded after the encoding area is determined by the finder/alignment patterns.

For \emph{thresholding}, we define a thresholding function $\psi$ to convert the grayscale sampled pixels $Q^g_x$ into binary format $Q_x^b$.
\begin{equation}
\label{equation:Get_t}
Q^b_x=\mathrm{\psi }\left(Q^g_x ,Q^t_x\right)=
\left\{
\begin{aligned}
           1 \ ,  \quad& \mathrm{if} \ Q^g_x \in [ \ Q^t_x,255 \ ]\\
           0 \ ,  \quad& \mathrm{if} \ Q^g_x \in [ \ 0 \ , \ Q^t_x \ )
\end{aligned}
\right.,
\end{equation}
where $Q^g_x$, $Q^t_x$, $Q^b_x$ are gray value, threshold, and binary results of pixel $x$, respectively. According to ZXing, the threshold $Q^t_x$ is not a constant and computed by a \emph{mean block binarization method} proposed in [4],[23].

\begin{figure}
\centering
\includegraphics[width=3.1in]{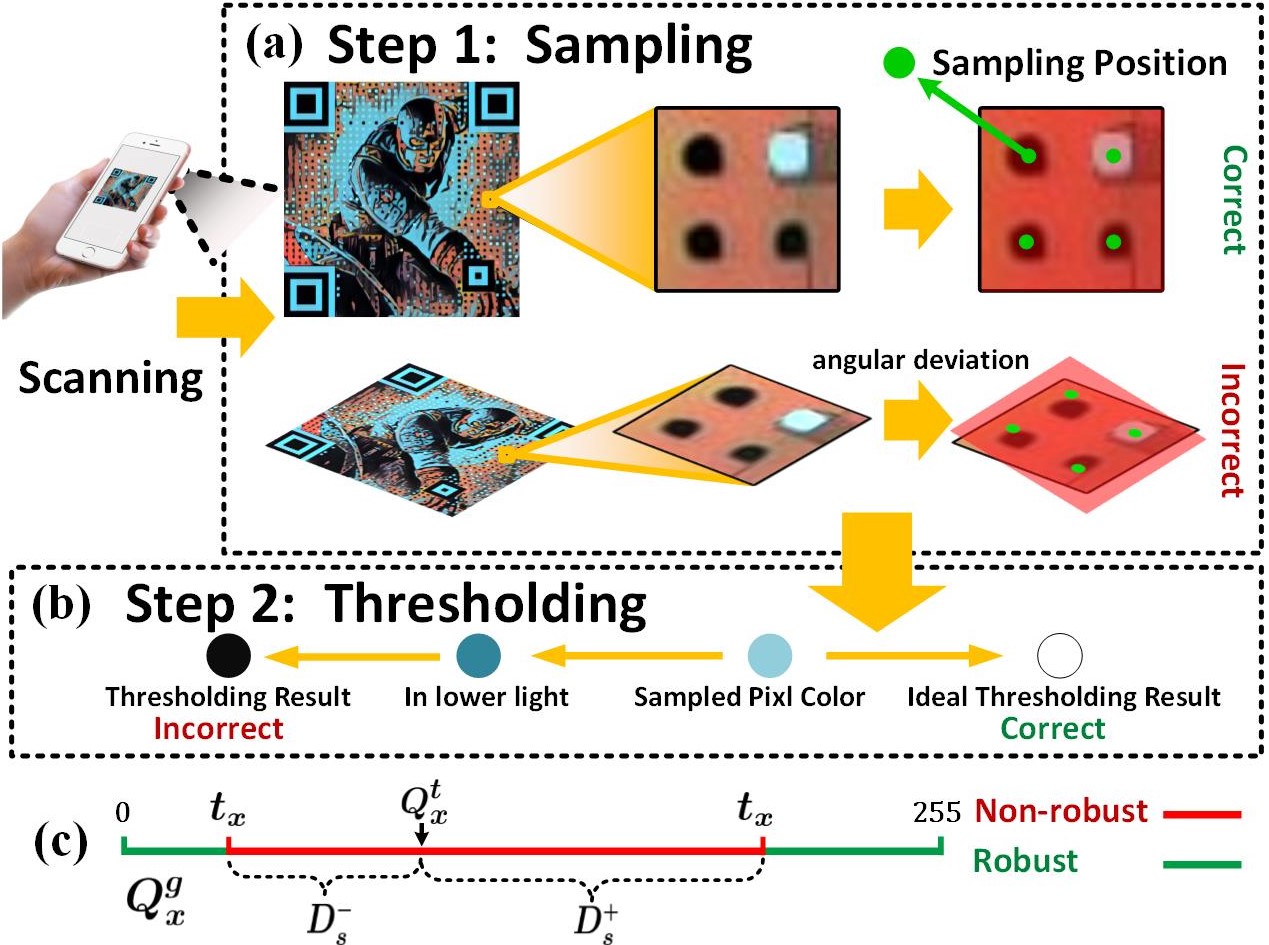}
\caption{(a) For sampling step, collected pixels may incorrect by external factors, e.g., image zoom, angle tilt, poor camera resolution. (b) For thresholding step, sampled pixels may be thresholded incorrectly by external factors, e.g., brightness, light's color. (c) To evaluate the robustness of pixel $x$, we set a finite interval, called non-robust region, in both positive and negative directions of $Q^t_{x}$ on the gray value axis.  }
\label{fig:4PC-sampel-thresholding}
\end{figure}
\begin{figure*}[t]
\centering
\includegraphics[width=5.2in]{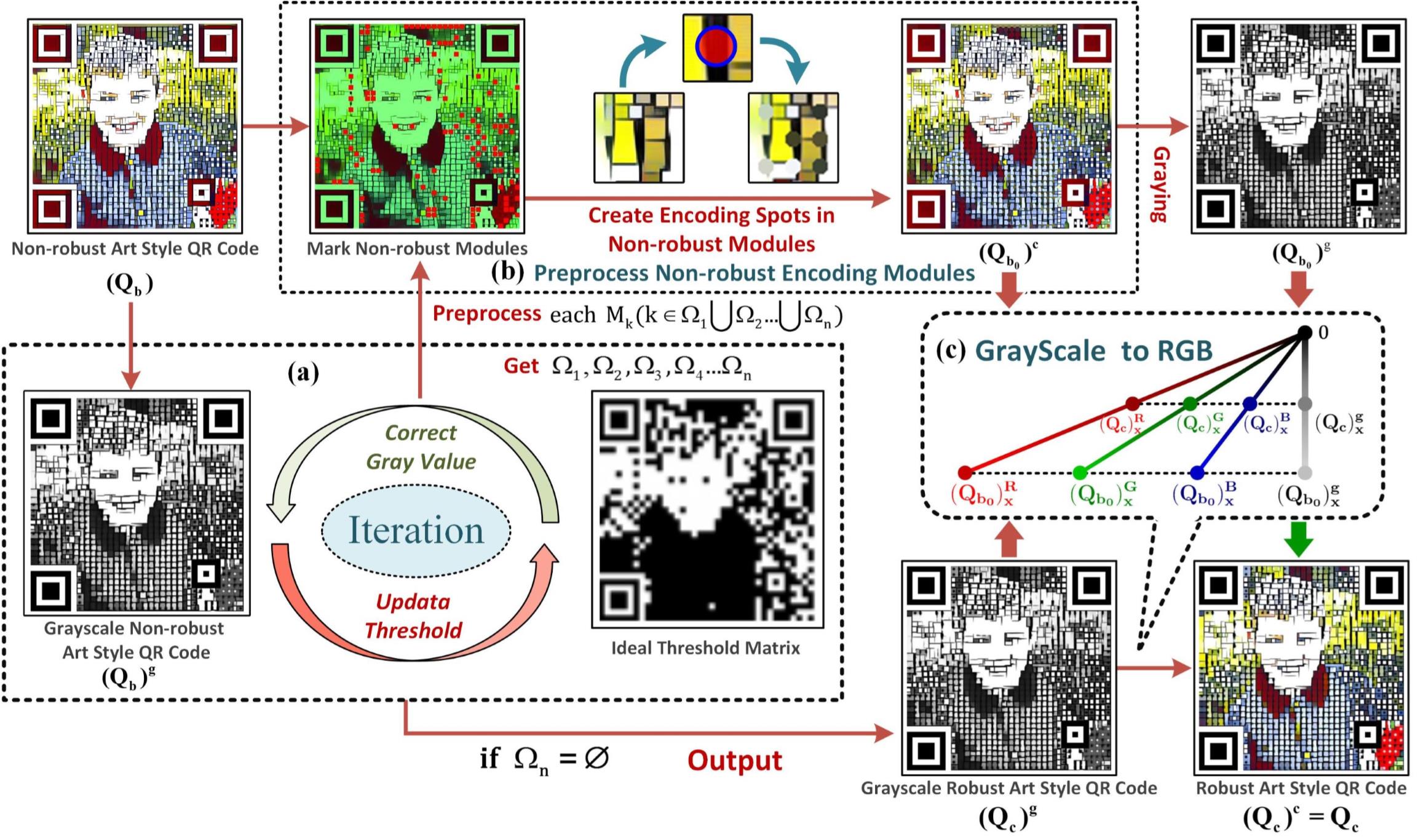}
\caption{Flowchart of Stage C. (a) Correcting non-robust modules, and updating thresholds, iteratively, until each module is robust. (b) Preprocessing non-robust modules by creating encoding spots. (c) Transforming grayscale robust art style QR code to RGB color and produce $Q_c$ finally. }
\label{fig:4PC}
\end{figure*}
\subsection{Robustness Evaluation of Encoding Modules}
In this subsection, we introduce how to estimate the robustness of modules in our error-correction mechanism. Following the two steps aforementioned, we evaluate the system robustness on sampling and thresholding. In reality, the translated message may differ from the ideal, due to the external factors (for sampling, e.g., image zoom, angle tilt, poor camera resolution; for thresholding, e.g., brightness, light's color). We can reduce the detriments of these factors by adapting the controllable attributes of QR codes, i.e., the modules' sizes and the modules' colors for optimizing the sampling (cf. Fig.~\ref{fig:4PC-sampel-thresholding}(a)) and thresholding (cf. Fig.~\ref{fig:4PC-sampel-thresholding}(b)) steps respectively.

For the robustness of sampling, Chu \emph{et al.} \cite{HF} proved that decoding a module of size $a\times a$ correctly requires an area at least $\frac{1}{3}a\times \frac{1}{3}a$ size in module's center contains the correct information. Intuitively, for the $k$-th module $M_k$ of the target QR code, in our setting, we employ a circular spot $S_{(k,\frac{1}{4}a)}$ of radius $\frac{1}{4}a$ pixels concentric with $M_k$ as a unit carrying the encoding messages. Our motivation to employ the encoding unit $S_{(k,\frac{1}{4}a)}$ is that the circular spot ensures the sampled pixels are the same in various scanning angles. Meanwhile, the size of $S_{(k,\frac{1}{4}a)}$ is larger than that of the theoretical valid size of module aforementioned in [7]. In addition, considering the dynamic requirements of aesthetic quality and robustness, we offer a configurable radius of the spot for users (cf.~Fig.~\ref{fig:4PC-0-1-delta}).

For the robustness of thresholding, given a QR code $Q$, the sampled threshold is distributed around the optimal threshold $ {Q}_x^{_{\s t}}$ due to the extraneous factors such as brightness and light color.  In fact, the closer ${Q}_x^g$  to the ideal boundary (0 or 255), the higher probability of ${Q}_x^g$ thresholding correctly (cf. Fig. \ref{fig:4PC-sampel-thresholding}(c)).
Accordingly, we next describe how to evaluate the robustness of module $M_k$ in $Q$. ZXing shows only the central pixels of modules influence the sampling results, which indicates the pixels closer to the center are more important. Hence, we define function $R_{M_k}$ by \emph{Gauss weight function} $G_{M_k}(x)$ to compute the robustness of $M_k$,
\begin{equation}
R_{M_k}=\sum_{x\in M_k} \xi(Q,x) \cdot G_{M_k}(x),
\end{equation}
where $G_{M_k}(x)$ is same as Eq.(\ref{equation:PA-Gauss}). $\xi(Q, x)$ is used to evaluate whether the pixel $x$ satisfies the robustness requirement under the constraint parameter $\delta$,
\begin{equation}
\label{equation:PC-Y}
\xi(Q,x)= 1-  \left[Q_x^i \oplus \psi (Q_x^g,t_x)\right],
\end{equation}
and $t_x$ is calculated by
\vspace{0.3cm}
\begin{equation}
\label{equation:PC-tx}
t_x=\left\{
\begin{aligned}
{Q}_x^t+D_{s}^{\sss +} \ , \ \ \mathrm{if} \ Q_x^i =1
\\{Q}_x^t-D_{s}^{\sss -} \ , \ \ \mathrm{if} \ Q_x^i =0
\end{aligned}
\right.,
\vspace{0.3cm}
\end{equation}
where ${Q}_x^i$ is the ideal thresholding result of ${Q_x^g}$, $D_{s}^{\sss -}=\delta \ | {Q}_{x}^t |$ and $D_{s}^{\sss +} =\delta \ |255-{Q}_x^t |$ represent non-robust region in the positive/negative direction respectively (cf. Fig. \ref{fig:4PC-sampel-thresholding}(c)). $Q_x^i$ is obtained by
\vspace{0.3cm}
\begin{equation}
\label{equation:PC-Qi}
Q_x^i=\left\{
\begin{aligned}
    &(Q_s)_x^b \ , \ \ \mathrm{if} \ x\in \{ S_{(1,r)}, S_{(2,r)}...S_{(\infty,r)}\}
\\  & \ \  Q_x^b \ \  \, , \ \ \ \! \mathrm{otherwise}
\end{aligned}
\right.,
\vspace{0.3cm}
\end{equation}
where $(Q_{s})_x^b$ and ${Q}_x^b$ are the thresholding results of pixel $x$ in standard QR Code ${Q_{s}}$ and ${Q}$ respectively.
Finally, $R_{M_k}\geqslant \eta$ or $R_{M_k}<\eta$ means $M_k$ is classified as a robust or non-robust module. The value of $\eta$ is set to 0.8 empirically.
\begin{figure}[t]
	\centering
	\includegraphics[width=2.8 in]{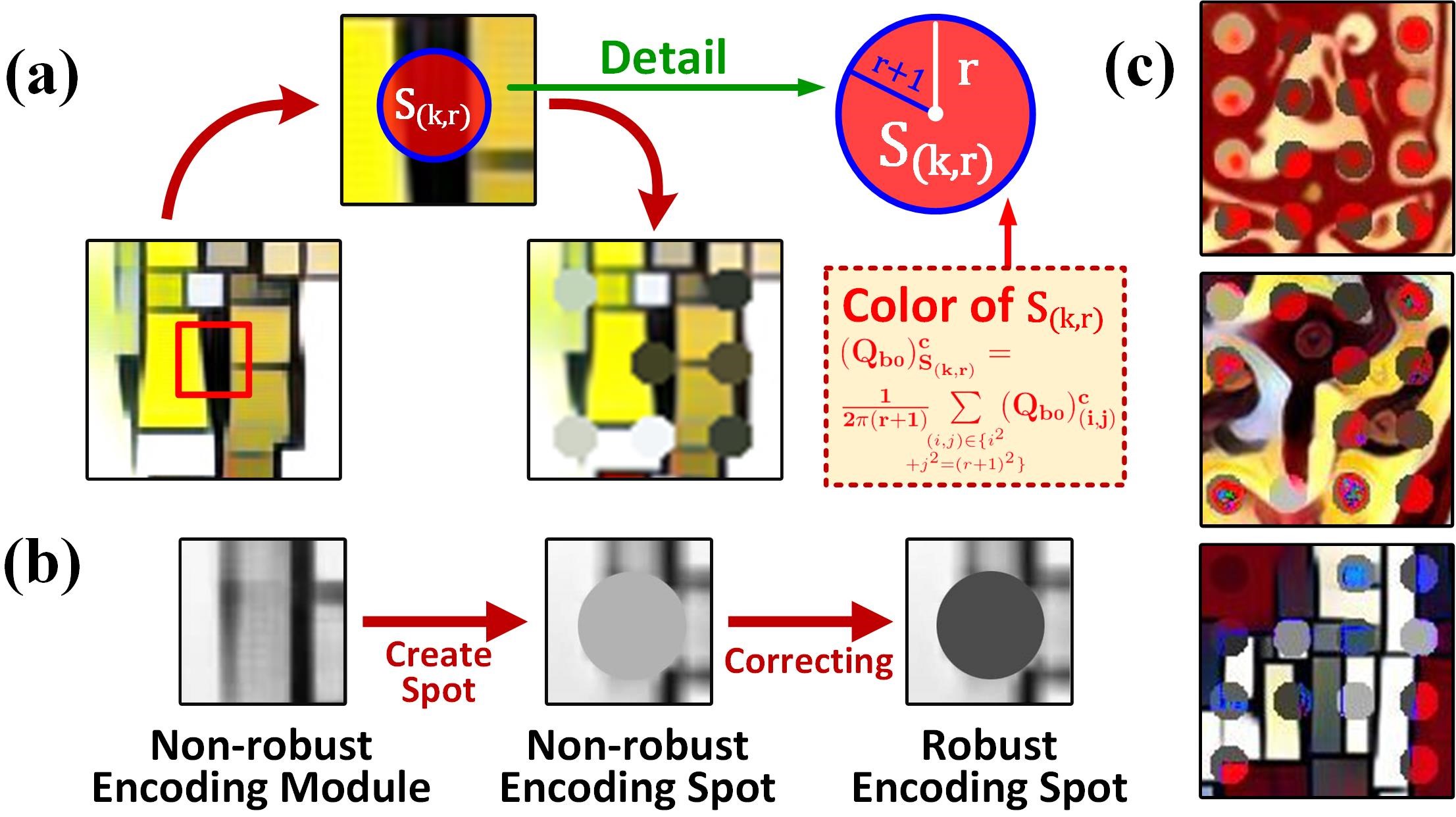}
	\caption{(a) Specific steps of Fig. \ref{fig:4PC}(b) that preprocess non-robust modules in $Q_b$, where the color of $S_{(k,r)}$ is computed by Eq.(\ref{eq:4PC-B0-color}). (b) Specific steps that correcting a grayscale non-robust encoding module. (c) The output $Q_c$ without processing by (a), i.e., the corrected spots have serious visual-unpleasant artifacts.}
	\label{fig:4PC-set_Ck}
\end{figure}
\subsection{Error-correction mechanism}
As illustrated in Fig. \ref{fig:4PC-sampel-thresholding}, our error-correction mechanism contains three main components: i) preprocessing non-robust modules in $Q_b$; ii) iterative-update based error correction; iii) transforming grayscale to RGB color.
\subsubsection{Preprocessing non-robust modules in $Q_b$} A serious issue is that the corrected encoding spots in $Q_c$ may incur visual-unpleasant artifacts during the process of correcting $Q_b$ (cf. Fig.12(c)). Aiming at solving this problem, we preprocess $Q_b$ to obtain $Q_{b{\sss 0}}$ via constructing spot in each non-robust module of $Q_b$ (cf. Fig.11(b) and Fig.12(a)). The color of each encoding spot $S_{(k,r)}$ in non-robust module $M_k$ is computed by
\begin{equation}
\label{eq:4PC-B0-color}
(Q_{b{\scriptscriptstyle 0}})_{S_{(k,r)}}^c =\textstyle{\frac{1}{2\pi(r+1)}} \!\!\!\!\!\! \sum\limits_{\substack{\scriptscriptstyle(i,j)\in\{i^2 \\ \scriptscriptstyle + j^2 =(r+1)^2\}}} \!\!\!\!\! (Q_{b{\scriptscriptstyle 0}})_{(i,j)}^c \ ,
\end{equation}
where $i, j$ are defined by Eq.(\ref{equation:PA-Gauss}), $Q_{b{\sss 0}}$ is utilized as the reference image that assists in transforming the grayscale $(Q_{c})^g$ to colored $(Q_c)^c$.
\subsubsection{Iterative-update based error correction}
As mentioned in \cite{Garateguy2014QR}, \cite{ISO}, the threshold of each pixel in QR code is computed by a \emph{mean block binarization method}, which means correcting a module may modify the thresholds of the adjacent pixels and incur additional error modules. Accordingly, in our error-correction mechanism, we correct non-robust modules and update thresholds iteratively until all modules are robust (cf. Fig.11(a)). The details of this mechanism are illustrated in Algorithm \ref{algorithm:1-all}.
\begin{algorithm}[h]
	\label{algorithm:1-all}
	\caption{Iterative update based overall flow}
	\LinesNumbered
	\KwIn{$Q_b\;, \: \delta \: $;}
	\KwOut{$Q_{b{\scriptscriptstyle 0}},(Q_c)^g;$}
	Initialize $n = 1,{\Omega} \neq \varnothing$\;
	$Q_{b{\scriptscriptstyle 0}} \leftarrow Q_{b},Q_{b{\scriptscriptstyle 1}} \leftarrow Q_{b} $\;
	\While {${\Omega}\neq \varnothing$}
	{ Detect non-robust modules in $Q_{b{\scriptscriptstyle 1}}$ by Eq.(11) \;
		\For {each non-robust module $M_k$ in $Q_{b{\scriptscriptstyle 1}}$}
		{Insert $k$ into $\text{\normalsize $\Omega $}$\;
			Create $S_{(k,r)}$ in the center of $M_k$ \;
			
			$(Q_{b{\scriptscriptstyle 1}})^{g\ast}_{x^{\vphantom{\prime}}} \leftarrow t_x$ ($x\in S_{(k,r)}$) \;
		}
		$(Q_{b{\scriptscriptstyle 1}})^{g} \leftarrow (Q_{b{\scriptscriptstyle 1}})^{g\ast}$ \;
		${\Omega}_n\leftarrow {\Omega}$\;
		$n\leftarrow n+1$\;
	}
	
	$({Q_{{c}}})^g \leftarrow (Q_{b{\scriptscriptstyle 1}})^g $\;
	\For {each $k\in \{{\Omega}_1 \bigcup {\Omega}_2 \cdots \bigcup {\Omega}_n \}$}
	{
		Create $S_{(k,r)}$ in the center of $M_k$ in $Q_{b{\scriptscriptstyle 0}}$\;
		\For{each $x\in S_{(k,r)}$}
		{   Compute $(Q_{b{\scriptscriptstyle 0}})_{(i,j)}^c$ by Eq.(\ref{eq:4PC-B0-color})\;
		}
	}
\end{algorithm}
\subsubsection{Transforming grayscale to RGB color}
After generating the robust grayscale art style QR code $(Q_{c})^g$, we further convert the grayscale image $(Q_{c})^g$ into the color image $(Q_{c})^c$ (cf. Fig.~\ref{fig:4PC-set_Ck}(c)). Let $\text{\Large $\zeta$}_{Q_x^c}=\left[\;Q^{ R}_x\;,\;Q^{ G}_x\;,\;Q^{B}_x \; \right]^T$ denote the color in RGB space of $Q_x^c$, where $Q^{ R}_x$, $Q^{ G}_x$, $Q^{B}_x$ represent the color values of pixel $x$ in R, G, B channels respectively. Here, we adopt a widely used formula
\begin{equation}
\label{equation:GRay}
{Q}_x^g=\alpha {Q}_x^{\sss R}+ \beta {Q}_x^{\sss G} +\gamma {Q}_x^{\sss B} \ ,
\end{equation}
 to calculate the grayscale value, where $\alpha=0.299$, $\beta =0.587$, $\gamma=0.114$, and $Q_{x}^g$ is the gray value of a pixel $x$ in grayscale QR code $Q^g$.
 We also construct a vector $\text{\Large $\kappa$}=(\alpha,\beta,\gamma)$ to deform  Eq.(\ref{equation:GRay}) into:
 \begin{equation}
 \label{equation:PC-RGB}
 {Q}^g_x=\text{\Large $\kappa$} \, \text{\Large $\zeta$}_{{Q}_x^c} \ .
 \end{equation}
Afterwards, let $\theta$ to denote the ratio of robust grayscale QR code $(Q_c)^g$ to the non-robust one $(Q_{b_0})^g$, in pixel $x$, as
 \begin{equation}
 \label{equation:PC-theta}
 \theta=\frac{(Q_{c})^g_{x^{\vphantom{t}} }}{(Q_{b{\sss 0}})_{x^{\vphantom{t}} }^g}=\frac{\text{\Large $\kappa$}\text{\Large $\zeta$}_{(Q_c)_x^c }} {\text{\Large $\kappa$}\text{\Large $\zeta$}_{(Q_{b{0}})_x^c} } \ .
 \end{equation}
 Combining Eq. (\ref{equation:PC-RGB}) and Eq. (\ref{equation:PC-theta}), we obtain
 \begin{equation}
 \text{\Large $\zeta$}_{  (Q_{c})_{x}^{c}}=\theta\text{\Large $\zeta$}_{(Q_{b0})^c_x } \ .
 \end{equation}
Therefore, we can get RGB color of each pixel in $Q_c$, meanwhile, $(Q_c)^c$ and $(Q_c)^g$ satisfy the conversion relation in Eq.(14), that is, $(Q_c)^c$ is as robust as $(Q_c)^g$.
\begin{figure}[t]
	\centering
	\includegraphics[width=3in]{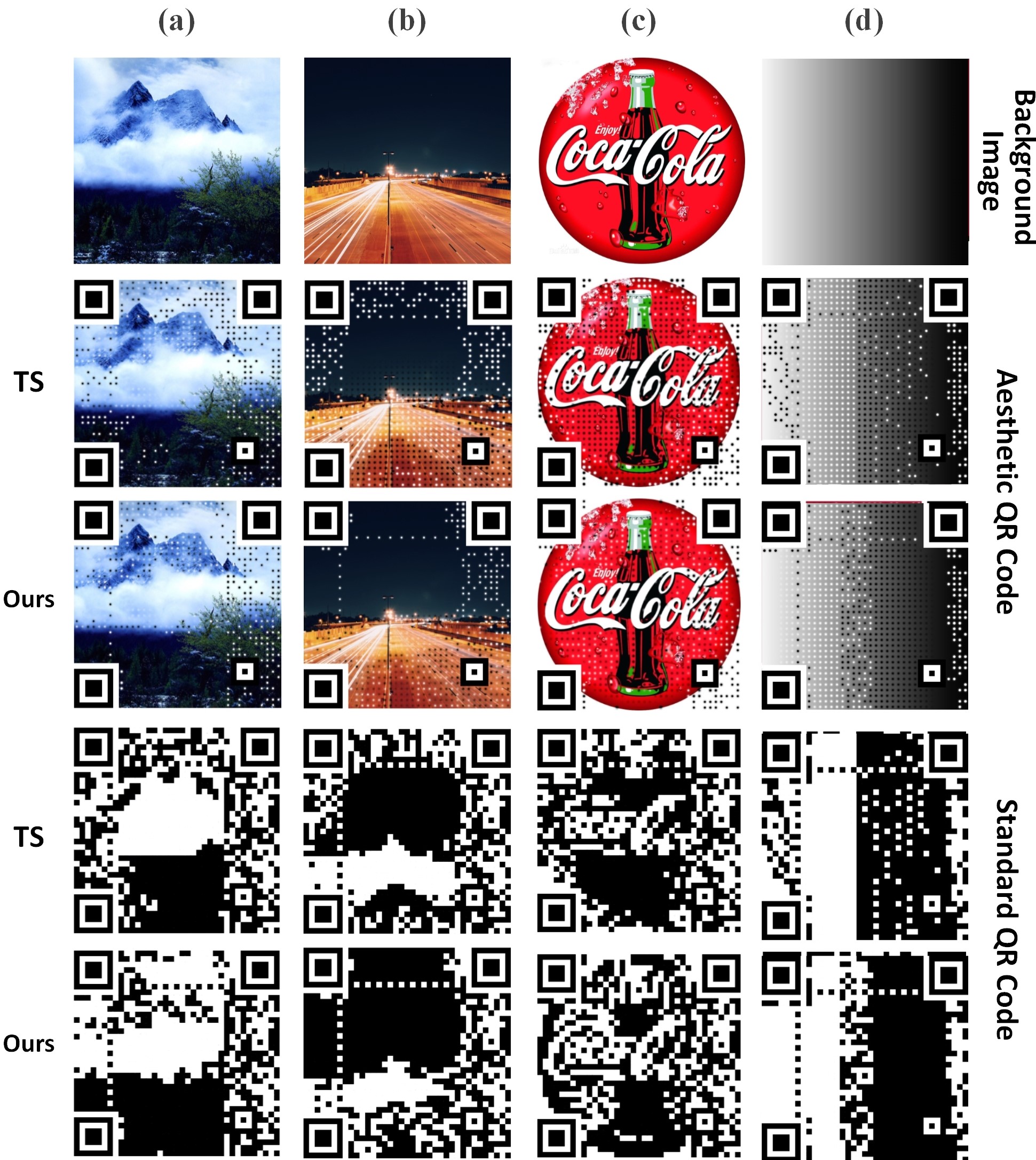}
	\caption{Comparison of experimental results between our $Q_a$ and TS (TS is the results of Two-Stage based method in [10]). Our results focus on the global feature of the blended image $I$, and the black/white encoding modules preferentially assigned to the darker/lighter color in $I$, which minimizes the visual contrast between noise-like modules and $I$. }
	\label{fig:3PA-example}
\end{figure}
\section{Experiment}
We conduct experiments on $Q_a$, $Q_b$, and $Q_c$ respectively. The experimental processes and results are described as follows three subsections.
\begin{figure*}[t]
	\centering
	\includegraphics[width=6in]{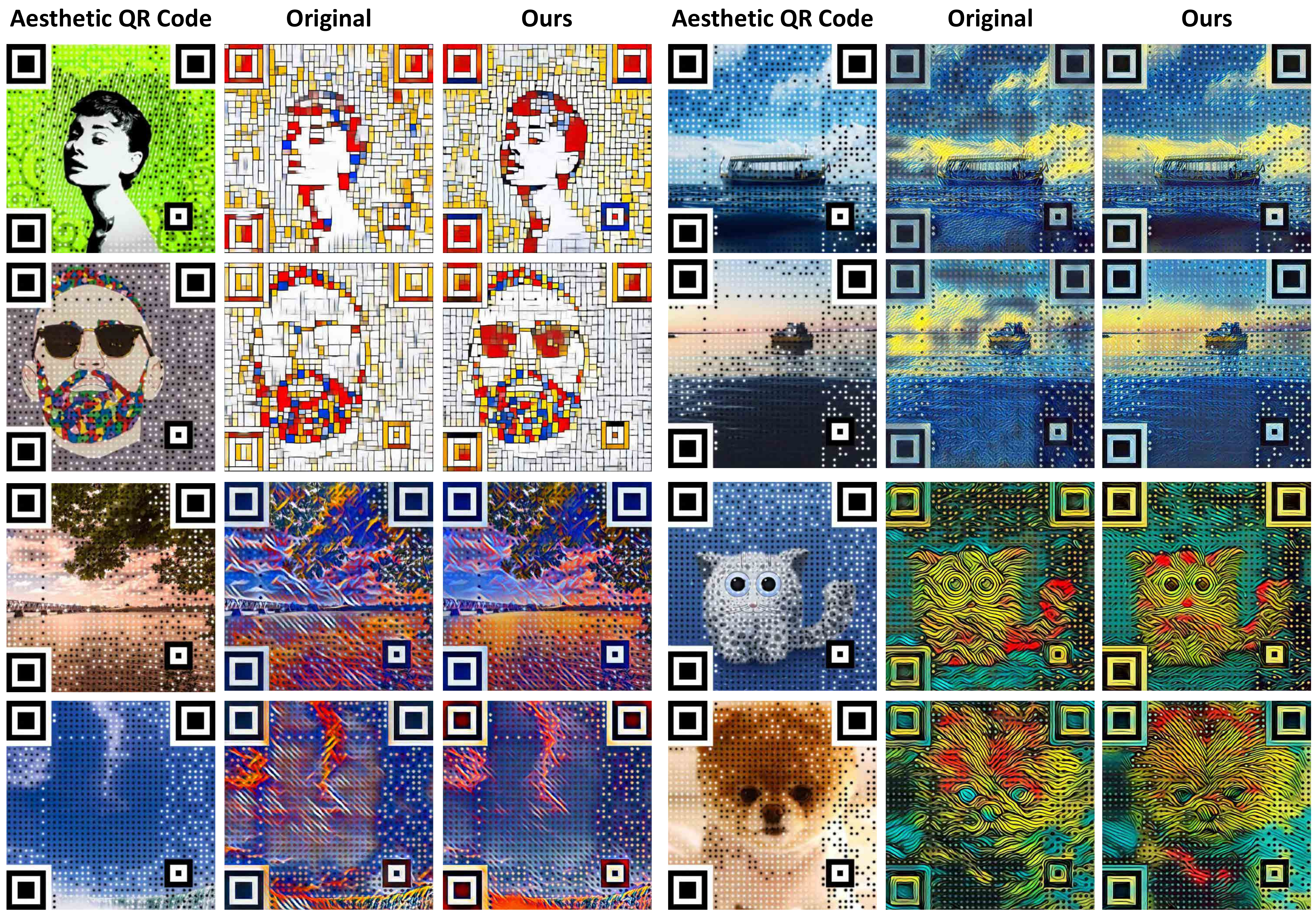}
	\caption{Results of original style transfer system \cite{Lff} and our refined one, which generated by the style target image indexed 4 (left upper), 6 (upper right), 9 (left lower), and 10 (right lower), in Fig. 15, respectively. In original results, irregular color changed occurs in some large regions, which extraordinary affects the visual quality and incurs encoding messages loss. Our refinement works well in robustness and visual quality. }
	\label{fig:3PB-GEcengduibi}
\end{figure*}
\begin{figure}[t]
	\centering
	\includegraphics[width=3.5in]{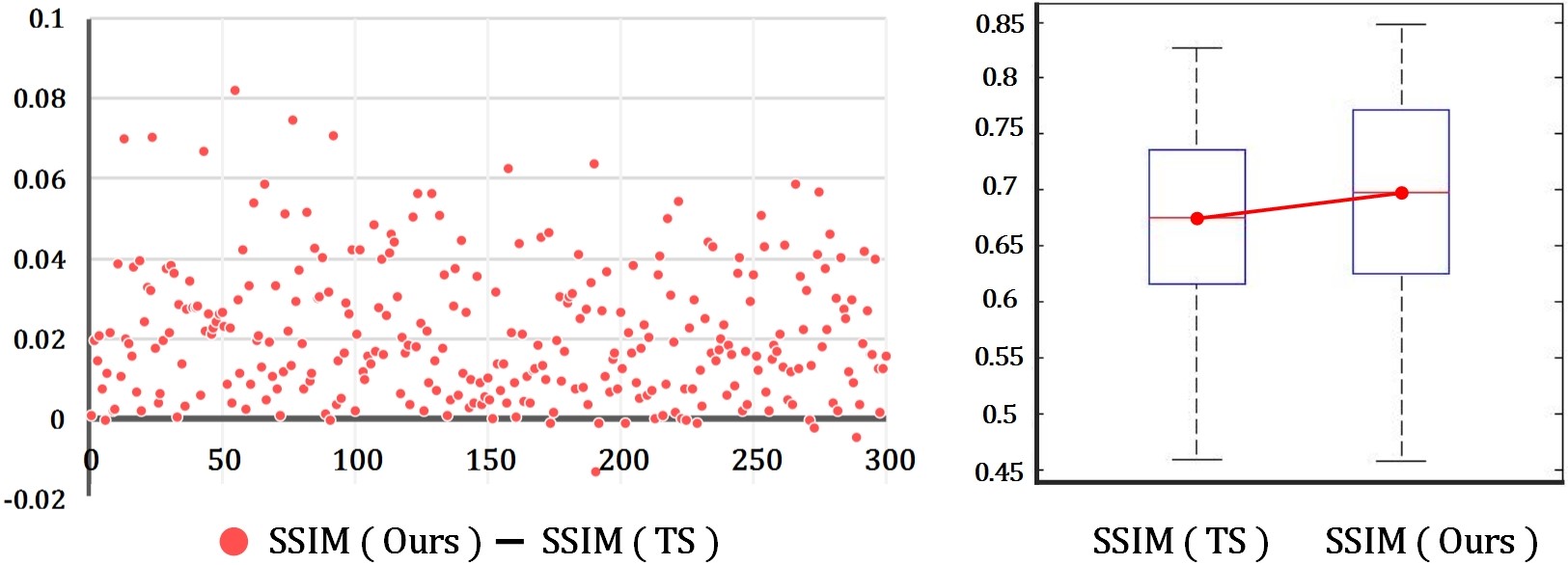}
	\caption{We experiment on indexed 1 to 300 blended images in dataset D, and produce baseline aesthetic QR codes by method \emph{TS}[10] and ours respectively. Left: The scatter-plot shows the resultant values of \emph{SSIM(Ours)} \emph{subtracts} \emph{SSIM(TS)}. Right: The box-plot shows the results of \emph{SSIM({Ours})} and \emph{SSIM(TS)}. The experimental results illustrate that our method outperforms \emph{TS} in the visual effect.}
	\label{fig:3PA-example-SSIM}
\end{figure}
\subsection{Experiments on $Q_a$}
\subsubsection{Experimental configuration of $Q_a$}
To evaluate the performance of TS \cite{TS} and our method, we prepare a dataset $D$ which contains 300 images of 512$\times$512 pixels with various contents (e.g., landscapes, cartoons, animals, characters, and trademarks). All images in $D$ are indexed from 1 to 300 and used as blended images for generating baseline aesthetic QR codes $Q_a$.

\subsubsection{Comparison of structure similarity}
We adopt a \emph{Structural SIMilarity (SSIM)} index proposed in~\cite{ssim} to measure the similarity between two images. Let \emph{SSIM(M)} denote the \emph{SSIM} index between the produced QR code and the corresponding blended image by \emph{M} (\emph{M} denotes TS or our method). Here, \emph{SSIM(M)} ranges from -1 to 1, \emph{SSIM(M)} = 1 means the aesthetic QR code is the same as the blended image.
As shown in Fig. \ref{fig:3PA-example}, in our results, the black/white modules are preferentially assigned to the locations with darkest/lightest color of the blended images, which effectively reduces the visual noise. Fig.~\ref{fig:3PA-example-SSIM} shows that the results of \emph{SSIM(Ours)} \emph{subtracts} \emph{SSIM(TS)} are positive in $96.3\%$ cases, which means our method outperforms \emph{TS} in the visual effect.

\begin{figure}[t]
	\centering
	\includegraphics[width=3.3in]{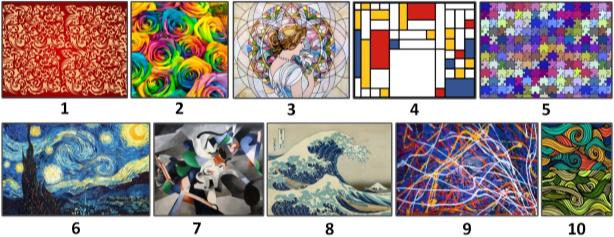}
	\caption{Style target images indexed from 1 to 10.}
	\label{fig:3PB-10zhongfengge}
\end{figure}

\subsection{Experiments on $Q_b$}
\subsubsection{Experimental configuration of $Q_b$}

We train the adapted style transfer network and the original one~\cite{Lff} on MS-COCO dataset \cite{MScoco}. Employing each of the two networks, we generate 300 $Q_{b}$ via combining 10 style target images (cf. Fig. \ref{fig:3PB-10zhongfengge}) and the 300 content target images $Q_a$ (output from the \emph{Experiment on $Q_a$ }) respectively.

\subsubsection{Comparisons of visual quality and robustness}  Fig. \ref{fig:3PB-GEcengduibi} shows the comparison of visual quality between the original style transfer system \cite{Lff} and our refined one. It can be found the baseline system used in \cite{Lff} excessively focuses on the high-level features, which is easily affected by the dense encoding modules and result in serious messy color blocks in outputs when $Q_a$ is as the content target. These messy color blocks significantly affect the visual quality and weaken the robustness of $Q_b$.

\begin{figure}[t]
	\centering
	\includegraphics[width=3.5in]{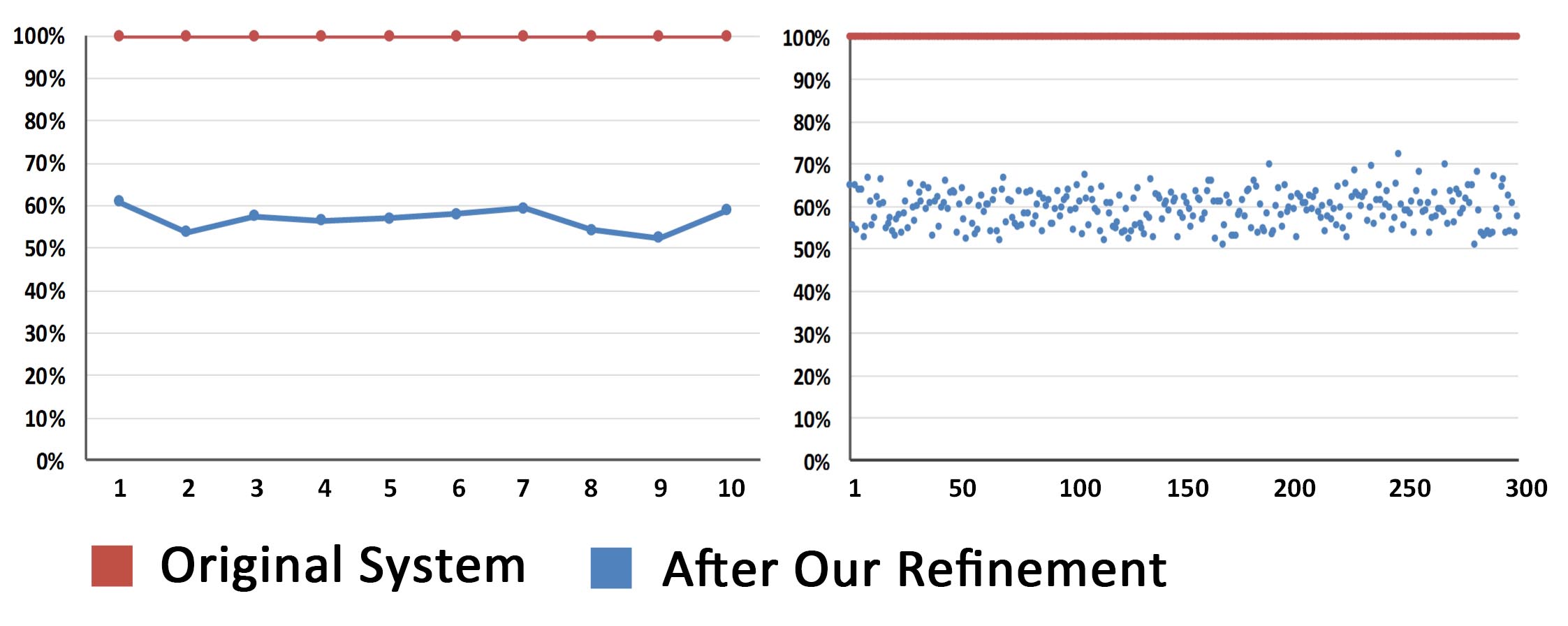}
	\caption{Left: The $100\%$ Stacked Line Chart displays the average numbers of error-modules in results, which produced by 10 styles (indexed from 1 to 10), and each of them combine with 300 content target images respectively (indicating the universal validity of each style). Right: The $100\%$ Stacked Scatter Chart, displays the average numbers of error-modules in results, which produced by 300 content target images (indexed from 1 to 300), and each of them combine with 10 styles respectively (indicating the universal validity of each content target). Experimental results show that the error-modules number in our result are approximate $\mathrm{60\%}$ of original system in both two situations, which means our refinement works well in robustness enhancing.}
	\label{fig:3PB-tongji}
\end{figure}

Aiming at evaluating the improvement on robustness, we calculate the average number of error modules in two situations: i) For 10 kinds of styles, each of them combines with 300 content target images respectively (evaluating the universal validity of each style); ii) For 300 content target images, each of them combines with 10 kinds of styles respectively (evaluating the universal validity of each content target). The experimental results as shown in Fig. \ref{fig:3PB-tongji}, the average error-modules number of our results are approximate $\mathrm{60\%}$ of the baseline system in both two situations.

To sum up, we make the style transformation system suitable for beautifying the baseline aesthetic QR code, which significantly improves the visual quality and robustness of $Q_b$.

\renewcommand\arraystretch{1.2}
\begin{table}[t]\footnotesize
	\caption{The meaning of each grade}
	\centering
	\begin{tabular}{p{1.5cm}<{\centering}|p{3.5cm}<{\centering}}
		\hline
		\hline
		\textbf{Grade} & \textbf{Indicate}   \\
		\hline
		5 &  very satisfied \\
		4 & satisfied \\
		3 & common \\
		2 & dissatisfied \\
		1 & very dissatisfied\\
		\hline
		\hline
	\end{tabular}
	\label{table:1Visual}
\end{table}
\renewcommand\arraystretch{1}

\subsection{Experiments on $Q_c$}
\renewcommand\arraystretch{1}
We evaluate the robustness and visual quality of $Q_c$ in following experiments.

\begin{figure}[t]
	\centering
	\includegraphics[width=2.8in]{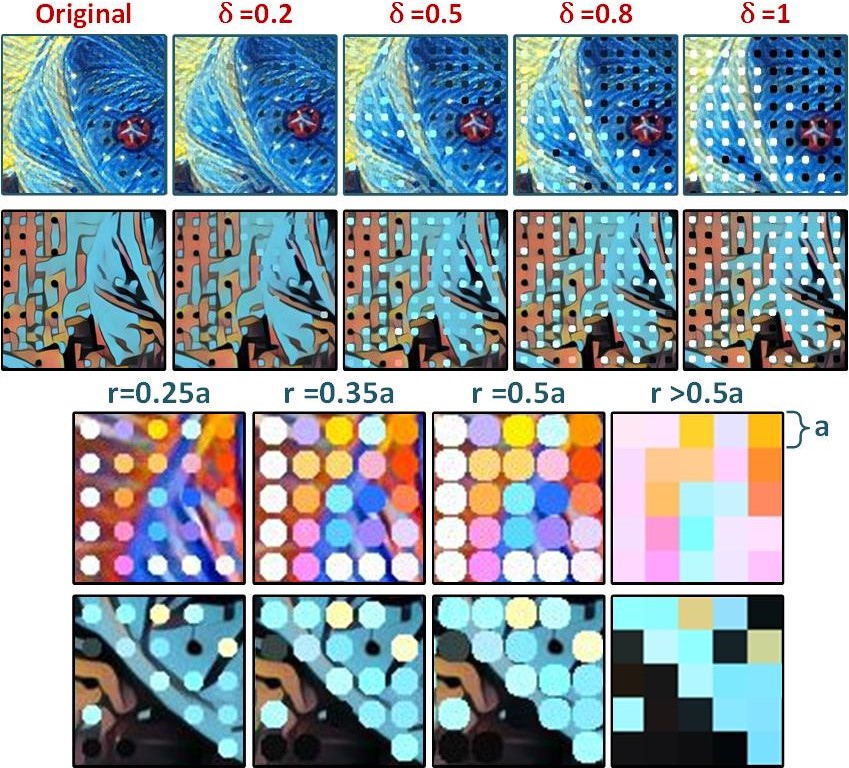}
	\caption{The appearance hanges of $Q_c$ with different $\delta$ and $r$. A larger $\delta$ or $r$ makes $Q_c$ more robust, and a smaller $\delta$ or $r$ makes $Q_c$ more similar to $Q_b$ that has higher visual quality.}
	\label{fig:4PC-0-1-delta}
\end{figure}
\begin{figure}[t]
	\centering
	\includegraphics[width=2.6in]{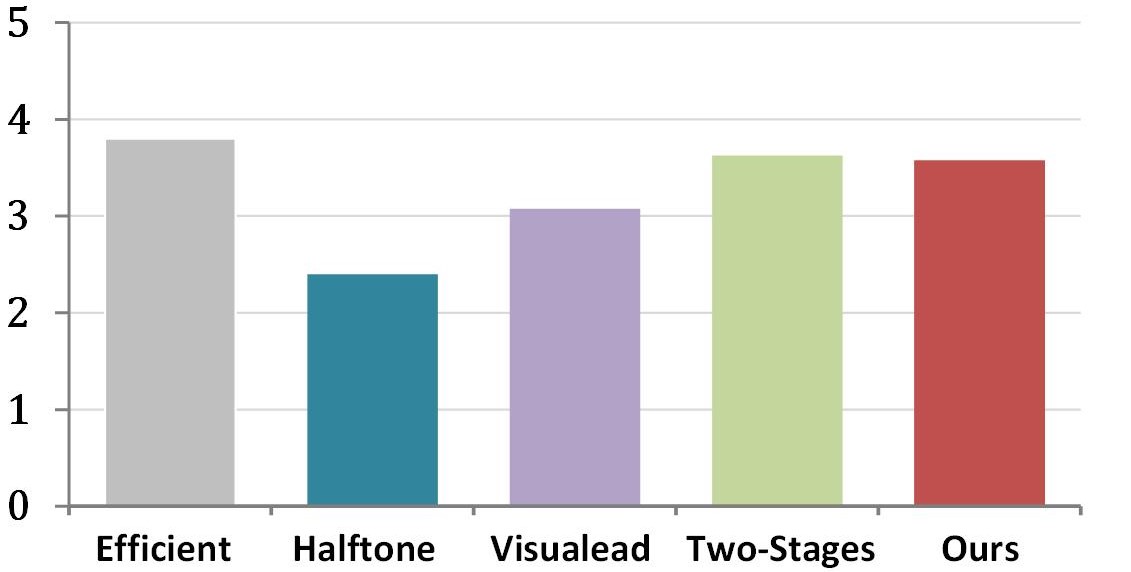}
	\caption{The result of the subjective test on attractiveness, which illustrates our SEE QR code reaches the state-of-the-art level in visual quality (we use the average score of 4 different styles as our final score).}
	\label{fig:Experiment-dafen2}
\end{figure}
\subsubsection{Influence of $\delta$ and radius $r$}
As mentioned in subsection B of section VI, varying parameters $\delta$ and $r$ is important for users to balance the performance of visual quality and robustness. As shown in the results~(c.f. Fig.~\ref{fig:4PC-0-1-delta}), a larger $\delta$ or $r$ incurs more noise-like encoding spots in $Q_c$, which reduces the visual quality yet enhances the robustness.

\renewcommand\arraystretch{1.1}
\begin{table}[t]\footnotesize
	\centering
	\caption{Decoding Rates on different Mobile Devices}
	\begin{tabular}{c|p{2cm}<{\centering}|p{0.8cm}<{\centering}|p{0.8cm}<{\centering}|p{0.8cm}<{\centering}}
		\hline
		\hline
		\multicolumn{1}{c|}{\multirow{2}{*}{\textbf{Moblie Phone}}}& \multirow{2}{*}{\textbf{App}}&\multicolumn{3}{p{3cm}<{\centering}}{\scriptsize{\textbf{Success Rate/Different Sizes}}}  \\ \cline{3-5}
		\multicolumn{1}{c|}{}                                &                      &\begin{minipage}{10cm}\vspace{0.1cm}
			{(3{cm})$^2$} \vspace{0.1cm}\end{minipage}     &\begin{minipage}{10cm}\vspace{0.1cm}
			{(5{cm})$^2$} \vspace{0.1cm}\end{minipage}      & \begin{minipage}{10cm}\vspace{0.1cm}
			{(7{cm})$^2$} \vspace{0.1cm}\end{minipage}     \\ \hline
		\multirow{4}{*}{Iphone 6s}                            & Wechat               &  $100\%$   &  $100\%$   &  $100\%$  \\ \cline{2-5}
		& Neo Reader           &  $100\%$   &  $100\%$   &  $100\%$  \\ \cline{2-5}
		& Alipay               &  $100\%$   &  $100\%$   &  $100\%$   \\ \cline{2-5}
		& QR Code Reader      &  $100\%$   &  $100\%$   &  $100\%$   \\ \hline
		\multicolumn{1}{c|}{\multirow{4}{*}{Huawei Honor 7}} & Wechat     &  $100\%$   &  $100\%$   &  $100\%$     \\ \cline{2-5}
		\multicolumn{1}{c|}{}                                & Neo Reader       &  $100\%$   &  $100\%$   &  $100\%$   \\ \cline{2-5}
		\multicolumn{1}{c|}{}                                & Alipay           &  $100\%$   &  $100\%$   &  $100\%$      \\ \cline{2-5}
		\multicolumn{1}{c|}{}                                & QR Code Reader     &  $100\%$   &  $100\%$   &  $100\%$    \\ \hline
		\multirow{4}{*}{Samsung Note 8}                       & Wechat             &  $100\%$   &  $100\%$   &  $100\%$     \\ \cline{2-5}
		& Neo Reader      &  $98\%$   &  $100\%$   &  $100\%$      \\ \cline{2-5}
		& Alipay          &  $100\%$   &  $100\%$   &  $100\%$       \\ \cline{2-5}
		& QR Code Reader   &  $100\%$   &  $100\%$   &  $100\%$     \\ \hline
		\multirow{4}{*}{Xiaomi Note 3}                        & Wechat           &  $100\%$   &  $100\%$   &  $100\%$      \\ \cline{2-5}
		& Neo Reader      &  $96\%$   &  $100\%$   &  $100\%$      \\ \cline{2-5}
		& Alipay          &  $100\%$   &  $100\%$   &  $100\%$       \\ \cline{2-5}
		& QR Code Reader    &  $100\%$   &  $100\%$   &  $100\%$      \\ \hline
		\hline
	\end{tabular}
	\label{table:scan-robust}
\end{table}
\subsubsection{Visual quality evaluation}
\begin{figure*}[htbp]
	\centering
	\includegraphics[width=6.4in]{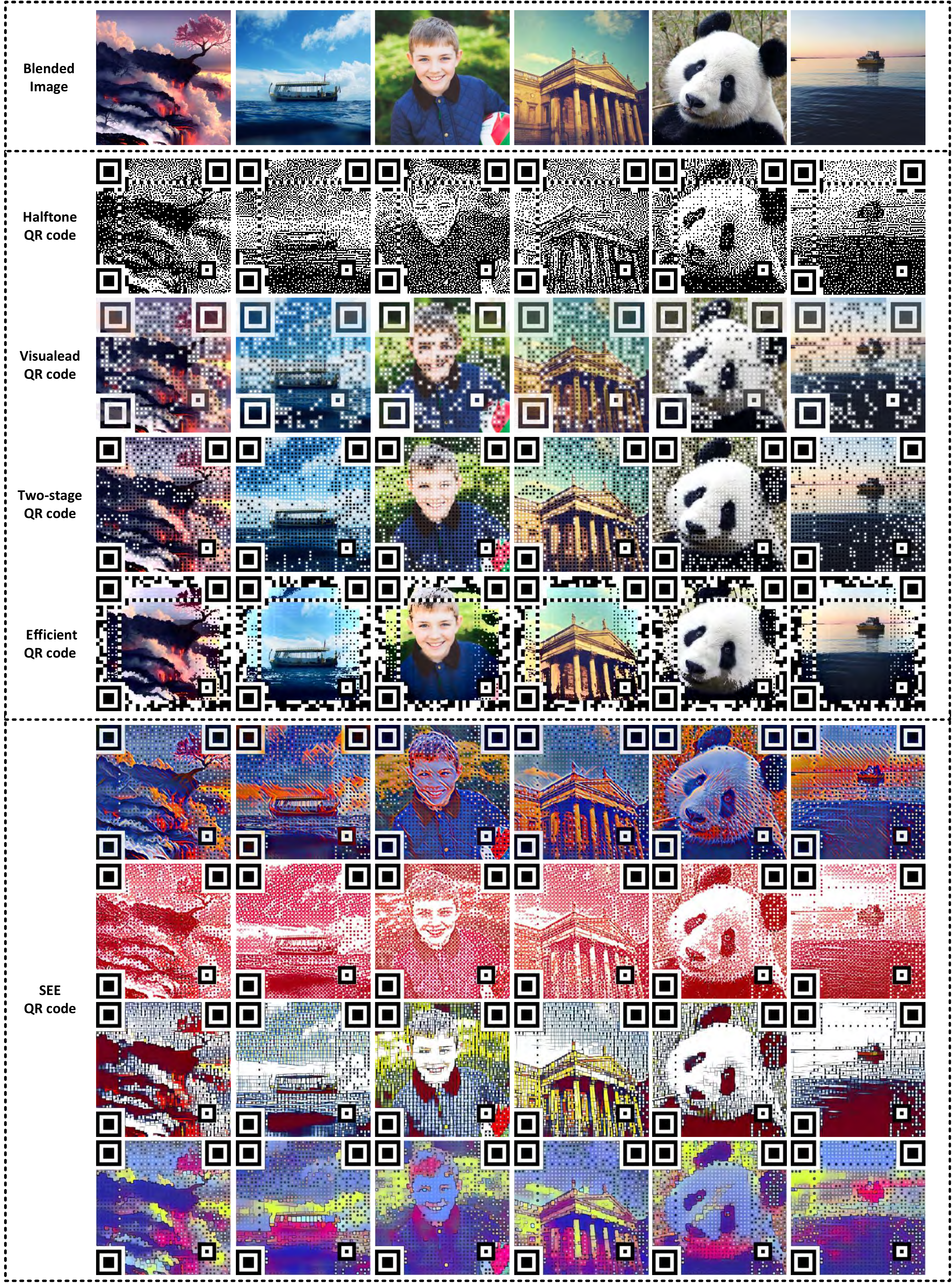}
	\caption{Some aesthetic QR codes examples generated by different existing methods and ours.}
	\label{fig:Experiment-dafen-example}
\end{figure*}

We conduct a user survey to evaluate the subjective visual quality. Preparing for the experiment, we randomly select 6 images from $D$ as the blended images to generated 6 group of QR codes. Each group includes 8 images: 4 SEE QR codes in different styles generated by us and 4 aesthetic QR codes of others (e.g., \emph{Visualead QR code}~\cite{VS}, \emph{Halftone QR code}~\cite{HF}, \emph{Efficent QR code}~\cite{EF}, and \emph{Two-Stages QR Code}~\cite{TS}). The produced QR codes corresponding to each method are presented in Fig.~\ref{fig:Experiment-dafen-example}, each image is of size $512\times512$ pixels and the version number of the QR code is 5.

We invited 40 volunteers (25 males and 15 females) irrelevant to this work to conduct a user study by scoring each group on a level of 1 to 5 (c.f. Table III), noted that our final score is the average score of 4 different styles. As shown in Fig.~\ref{fig:Experiment-dafen2}, compared with the state-of-the-art, our SEE QR code is more personalized and diversity without compromising the visual quality.

\subsubsection{Robustness evaluation}
According to the decoding principle of QR code, in ideal condition, $Q_c$ can be decoded correctly when $\delta>0$. However, as mentioned in Section IV, $Q_c$ may be unreadably by a camera, due to external factors, e.g., angle tilt, poor camera resolution, light color, brightness. Therefore, we design an experiment to evaluate the robustness of $Q_c$ in real scene.

 We examine the robustness on 4 group of our SEE QR code which used in the experiment of \emph{visual quality evaluation}. Meanwhile, considering the influence of image size on decoding performance, we scan each example with size 3cm$\times$3cm, 5cm$\times$5cm, and 7cm$\times$7cm, respectively, by different mobile phones and QR decoders. The successful decoding rate is calculated via $\frac{\mathrm{Successful \ decoding \ times}}{\mathrm{Scanning \ times}}\times100\%$ in 50 scanning times.

As shown in Table IV, the successful decoding rates are always greater than $96\%$, which means our SEE QR code is robust enough for daily applications. In addition, users can also enhance the resultant robustness by increasing $\delta$ or $r$ with sacrificing little visual quality.

\section{Conclusion}
In this paper, we propose a novel automatic approach equipping with a robust error correction mechanism to generate beautiful art style QR code called \emph{SEE QR code}. Compared with the state-of-the-art, our SEE QR code achieves better performance in the perspectives of personalization, artistry, and robustness, which can efficiently support the real-life application and business promotion.
\section*{Acknowledgement}
We are grateful to the anonymous reviewers for their comments and suggestions.
 The authors would like to thank J. Johnson, A. Alahi, and F.-F. Li for providing the source codes of their works, and also appreciate S.-S. Lin, M.-C. Hu, C.-H. Lee, and T.-Y. Lee help on providing example results of their work for evaluation, so the authors could easily compare their method with the state-of-the-art works.

\bibliographystyle{IEEEtran}%
\bibliography{IEEEabrv,my20180302}

\vspace{-1cm}
\begin{IEEEbiography}[{\includegraphics[width=1in,height=1.25in,clip,keepaspectratio]{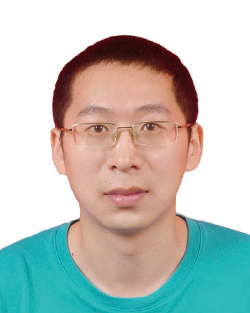}}]{Mingliang Xu}
\label{Authors_photos}
is an associate professor in the School of Information Engineering of Zhengzhou University, China, and currently is the director of CIISR ( Center for Interdisciplinary Information Science Research), and the general secretary of ACM SIGAI China. His research interests include virtual reality and artificial intelligence. Xu got his Ph.D. degree in computer science and technology from the State Key Lab of CAD\&CG at Zhejiang University.
\end{IEEEbiography}

\vspace{-1cm}
\begin{IEEEbiography}[{\includegraphics[width=1in,height=1.25in,clip,keepaspectratio]{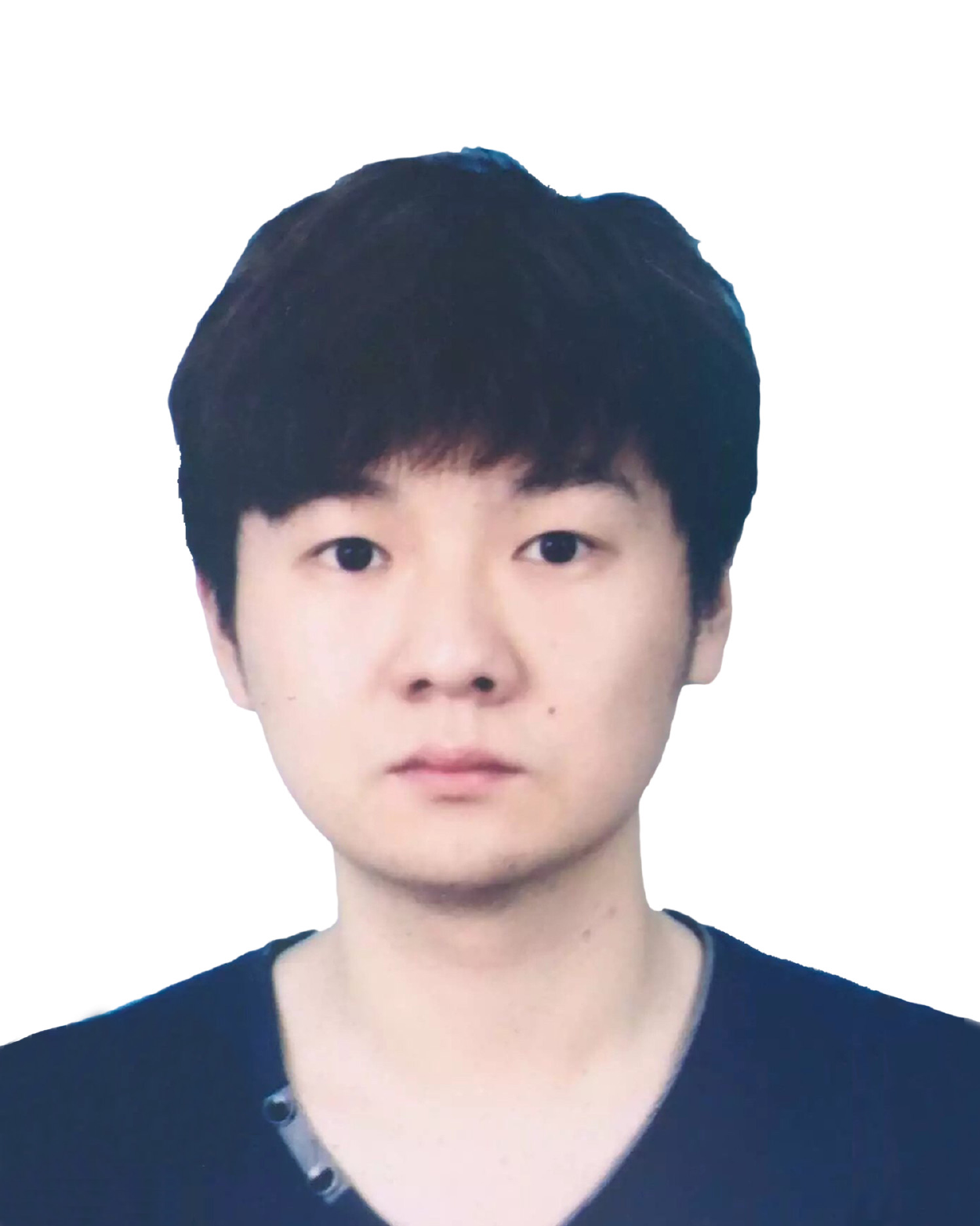}}]{Hao Su}
is a master student in Center for Interdisciplinary Information Science Research, Zhengzhou University, China. He received the B.E. degree in Computer Science and Technology from Zhengzhou University, in 2016. His research interests include computer version, image processing, and computer graphics.
\end{IEEEbiography}

\vspace{-1.1cm}
\begin{IEEEbiography}[{\includegraphics[width=1in,height=1.25in,clip,keepaspectratio]{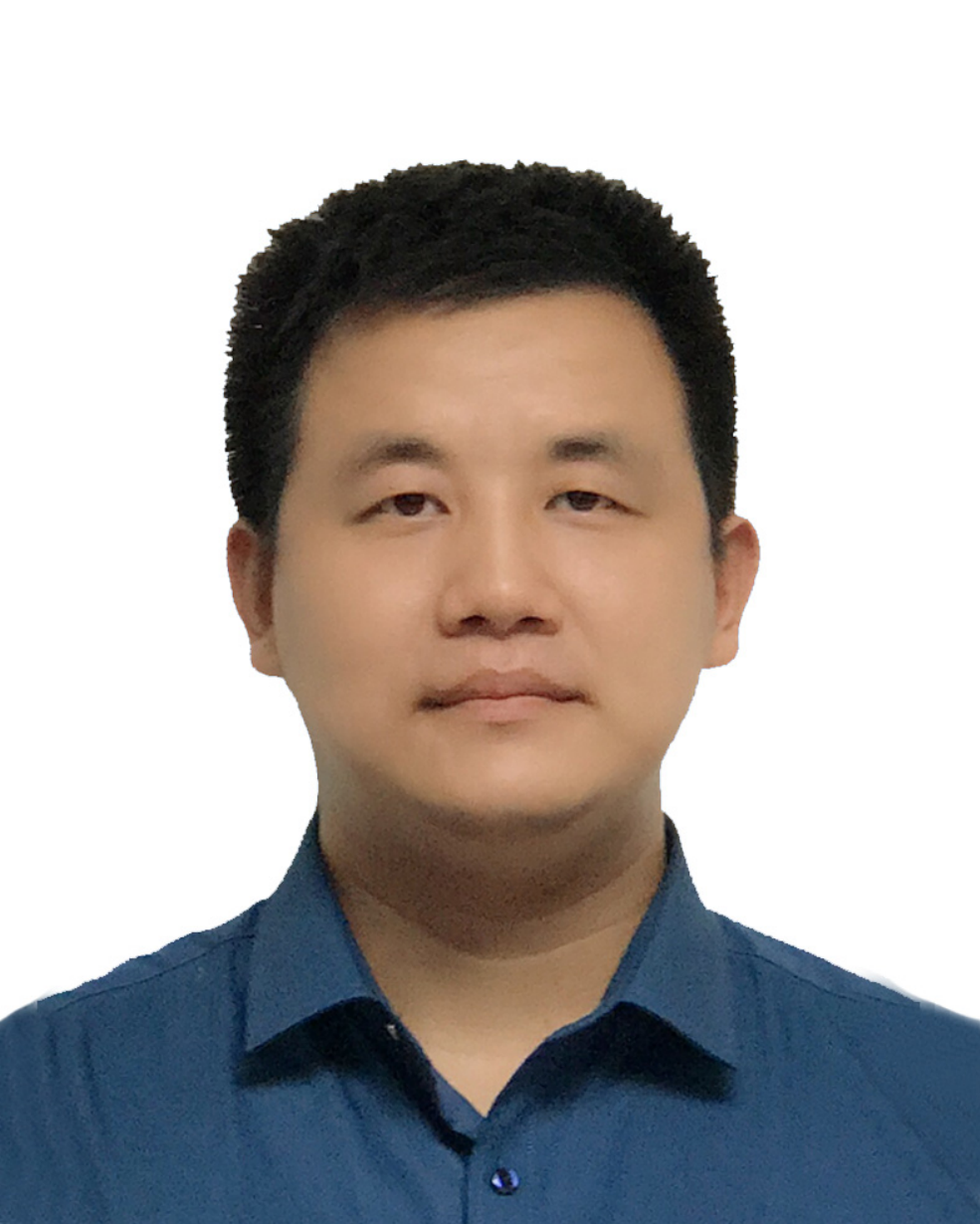}}]{Yafei Li}
received the PhD degree in computer science from Hong Kong Baptist University, in 2015. He is an assistant professor in the School of Information Engineering, Zhengzhou University, Zhengzhou, China. He holds a visiting position in the Database Research Group (http:// www.comp.hkbu.edu.hk/db) with Hong Kong Baptist University. His research interests include mobile and spatial data management, locationbased services, and smart city computing
\end{IEEEbiography}

\vspace{-1.1cm}
\begin{IEEEbiography}[{\includegraphics[width=1in,height=1.25in,clip,keepaspectratio]{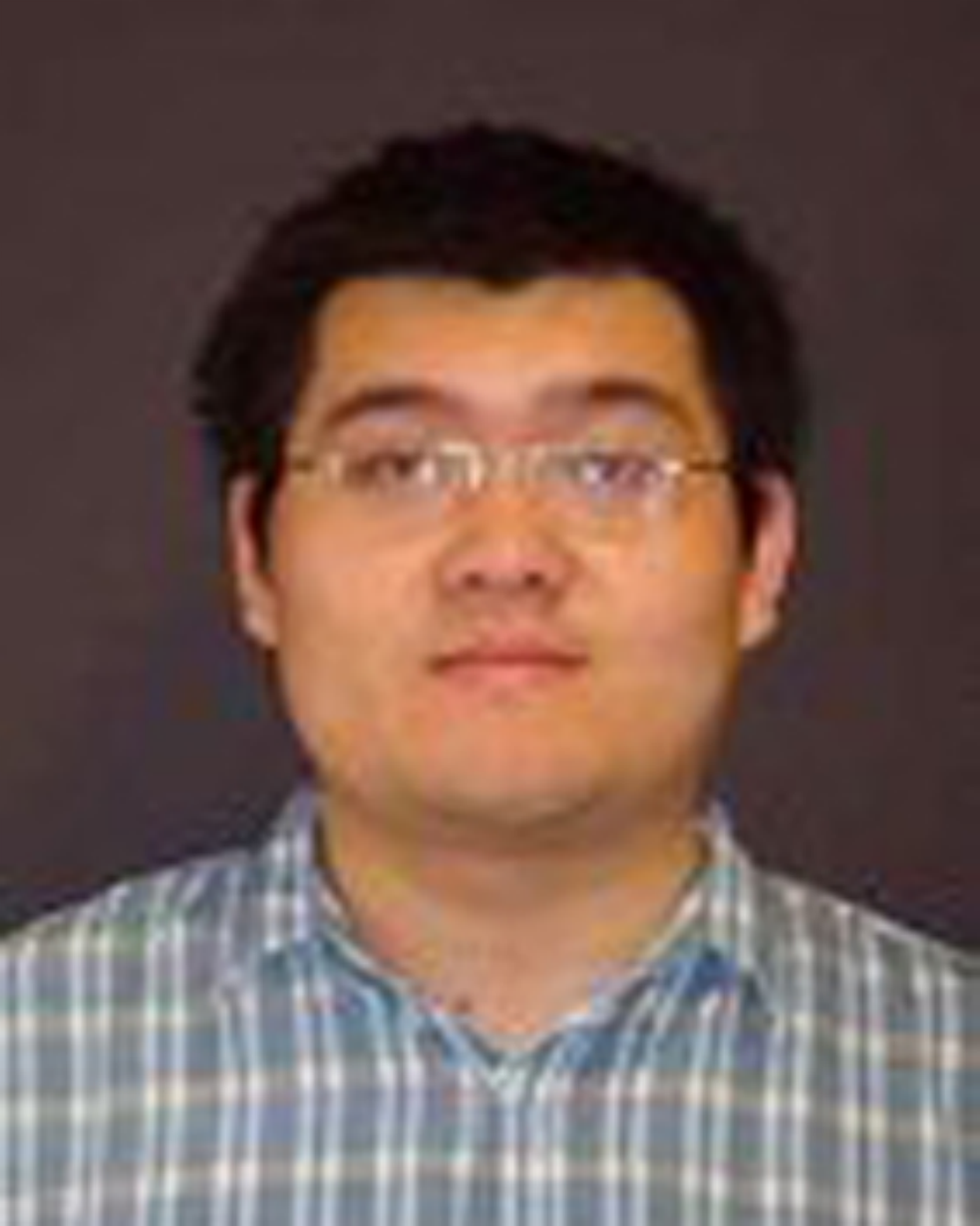}}]{Xi Li}
received the Ph.D. degree from the National Laboratory of Pattern Recognition, Chinese Academy of Sciences, Beijing, China, in 2009. From
2009 to 2010, he was a Post-Doctoral Researcher with CNRS Telecomd ParisTech, France. He was a Senior Researcher with the University of
Adelaide, Australia. He is currently a Full Professor with Zhejiang University, China. His research interests include visual tracking, motion analysis, face recognition, Web data mining, and image and video retrieval.is an associate professor .
\end{IEEEbiography}

\vspace{-1.1cm}
\begin{IEEEbiography}[{\includegraphics[width=1in,height=1.25in,clip,keepaspectratio]{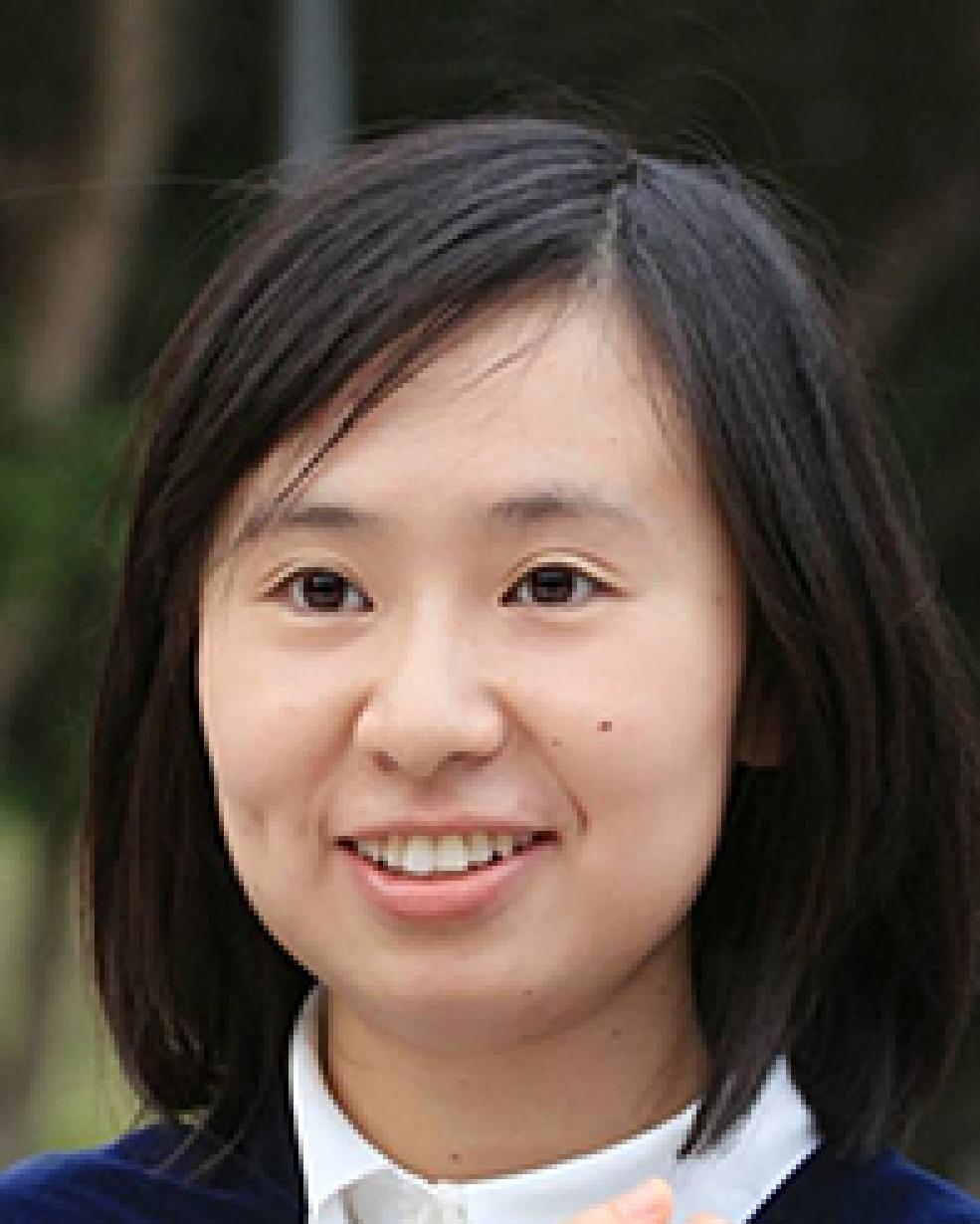}}]{Jing Liao}
received the dual Ph.D. degrees from Zhejiang University and Hong Kong University of Science and Technology in 2014 and 2015 respectively. She is currently a researcher in Visual Computing Group at Microsoft Research Asia (MSRA). Her research interests include image and video processing, computational photography, non-photo-realistic rendering.
\end{IEEEbiography}

\vspace{-1.1cm}
\begin{IEEEbiography}[{\includegraphics[width=1in,height=1.25in,clip,keepaspectratio]{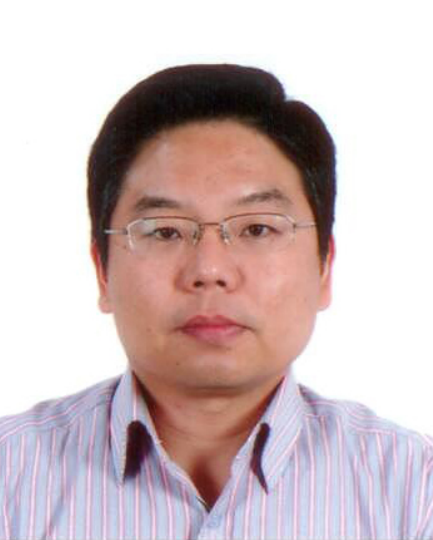}}]{Jianwei Niu}
received the M.S. and Ph.D. degrees in computer science from Beihang University, Beijing, China, in 1998 and 2002, respectively. He was a visiting scholar at School of Computer Science, Carnegie Mellon University, USA from Jan. 2010 to Feb. 2011. He is a professor in the School of Computer Science and Engineering, BUAA, and an IEEE senior member. His current research interests include mobile and pervasive computing, mobile video analysis.
\end{IEEEbiography}

\vspace{-1.1cm}
\begin{IEEEbiography}[{\includegraphics[width=1in,height=1.25in,clip,keepaspectratio]{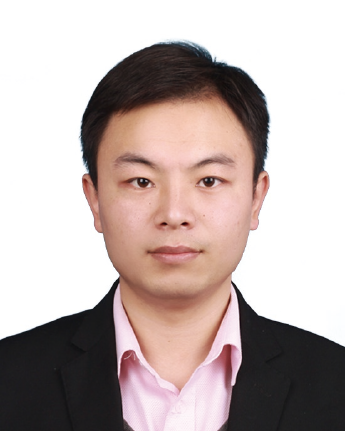}}]{Pei Lv}
is an assistant professor in Center for Interdisciplinary Information Science Research, Zhengzhou University, China.His research interests include video analysis and crowd simulation. He received his Ph.D in 2013 from the State Key Lab of CAD\&CG, Zhejiang University, China.
\end{IEEEbiography}

\vspace{-17cm}
\begin{IEEEbiography}[{\includegraphics[width=1in,height=1.25in,clip,keepaspectratio]{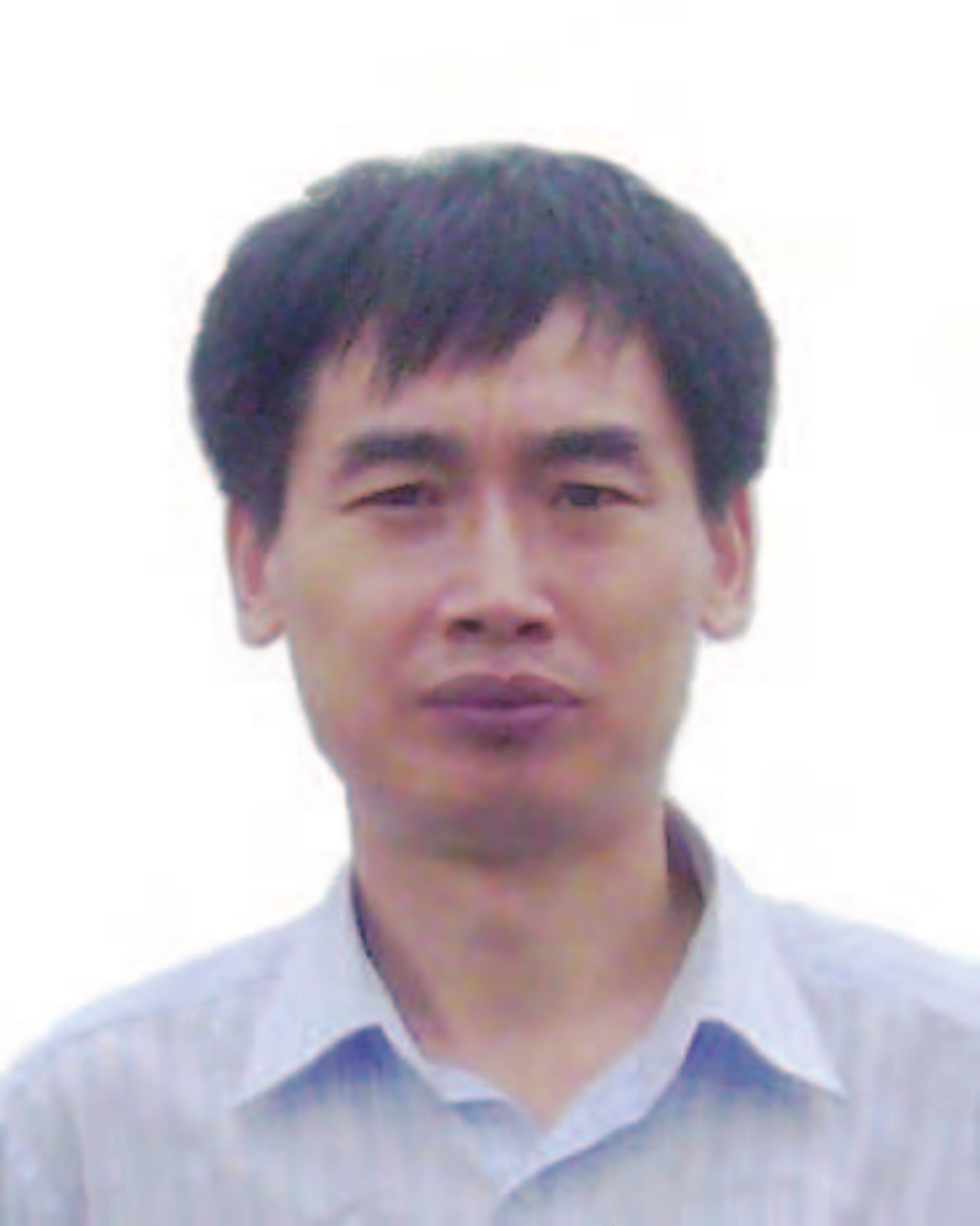}}]{Bing Zhou}
is currently a professor in Center for Interdisciplinary Information Science Research, Zhengzhou University, Henan, China. He received the B.S. and M.S. degrees from Xiâan Jiaotong University in 1986 and 1989, respectively,and the Ph.D. degree in Beihang University in 2003, all in computer science. His research interests cover video processing and understanding, surveillance, computer vision, multimedia applications.
\end{IEEEbiography}

\end{document}